\documentclass[showpacs,aps,prd,nofootinbib,floatfix,amsmath,amssymb,twocolumn]{revtex4}
\usepackage{graphicx}
\usepackage[dvipsnames]{xcolor}
\usepackage{dsfont}
\begin{document}

\makeatletter
\newbox\slashbox \setbox\slashbox=\hbox{$/$}
\newbox\Slashbox \setbox\Slashbox=\hbox{\large$/$}
\def\pFMslash#1{\setbox\@tempboxa=\hbox{$#1$}
  \@tempdima=0.5\wd\slashbox \advance\@tempdima 0.5\wd\@tempboxa
  \copy\slashbox \kern-\@tempdima \box\@tempboxa}
\def\pFMSlash#1{\setbox\@tempboxa=\hbox{$#1$}
  \@tempdima=0.5\wd\Slashbox \advance\@tempdima 0.5\wd\@tempboxa
  \copy\Slashbox \kern-\@tempdima \box\@tempboxa}
\def\FMslash{\protect\pFMslash}
\def\FMSlash{\protect\pFMSlash}
\def\miss#1{\ifmmode{/\mkern-11mu #1}\else{${/\mkern-11mu #1}$}\fi}
\makeatother

\title{Electric dipole moments of charged leptons at one loop in the presence of massive neutrinos}

\author{H. Novales-S\'anchez$^{(a)}$}
\author{M. Salinas$^{(a)}$}
\author{J. J. Toscano$ {}^{(a,b)}$}
\author{O. V\'azquez-Hern\'andez$^{(a)}$}
\affiliation{$^{(a)}$Facultad de Ciencias F\'{\i}sico Matem\'aticas,
Benem\'erita Universidad Aut\'onoma de Puebla, Apartado Postal
1152, Puebla, Puebla, M\'exico.\\
$^{(b)}$Facultad de Ciencias F\'isico Matem\' aticas, Universidad Michoacana de San Nicol\' as de Hidalgo, Avenida Francisco J. M\'ujica S/N,
58060, Morelia, Michoac\' an, M\' exico.}

\begin{abstract}
Violation of $CP$ invariance is a quite relevant phenomenon that is found in the Standard Model, though in small amounts. This has been an incentive to look for high-energy descriptions in which $CP$ violation is increased, thus enhancing effects that are suppressed in the Standard Model, such as the electric dipole moments of elementary particles. In the present investigation, we point out that charged currents in which axial couplings are different from vector couplings are able to produce one-loop contributions to electric dipole moments of charged leptons if neutrinos are massive and if these currents violate $CP$. We develop our discussion around charged currents involving heavy neutrinos and a $W'$ gauge boson coupling to Standard Model charged leptons. Using the most stringent bound on the electron electric dipole moment, provided by the ACME Collaboration, we determine that the upper bound on the difference between axial and vector currents lies within $\sim10^{-10}$ and $\sim10^{-7}$ for heavy-neutrino masses between $0.5\,{\rm TeV}$ and $6\,{\rm TeV}$ and if the $W'$ mass is within $0.45\,{\rm TeV}-7\,{\rm TeV}$. This possibility is analyzed altogether with the anomalous magnetic moments of charged leptons, among which we estimate, for the $\tau$ lepton, an anomalous magnetic moment contribution between $\sim10^{-8}$ and $\sim10^{-10}$ for neutrino masses ranging from $0.5\,{\rm TeV}$ to $6\,{\rm TeV}$ and a $W'$ mass between $0.45\,{\rm TeV}$ and $7\,{\rm TeV}$. The general charged currents are also used to calculate the branching ratio for $\mu\to e\gamma$, which gets suppressed if the set of masses of heavy neutrinos is quasidegenerate. In a scenario of nondegenerate neutrino masses, we find that regions of neutrino and $W'$ masses in which the contributions to this flavor changing branching ratio are lower than the current upper bound exist. We show that such regions can be widened if the $W'$ gauge boson mass is larger.
\end{abstract}


\maketitle
\section{Introduction}
\label{int}
The discovery of the Higgs-like particle~\cite{Higgs,EnBr} with mass around 125\,GeV, announced by the CMS~\cite{HiggsATLAS} and ATLAS~\cite{HiggsCMS} Collaborations at the Large Hadron Collider, has been a remarkable achievement, which, however, is not a statement that the Standard Model is the last part of the story. The premise that there is a more fundamental physical description, beyond the Standard Model, was fed during years by theoretical issues, but hints of its nature have finally been provided by experimental observations that include neutrino oscillations~\cite{SK1998,SNO2002}, dark matter~\cite{Zwicky,Rubin,DMlensing}, and perhaps even a new particle with mass $\sim$750\,GeV~\cite{A750,CMS750}.
The phenomenon of neutrino oscillations, first observed at Super-Kamiokande, then at the SNO, and recently confirmed by the determination of the last mixing angle by the Daya Bay~\cite{DB2012} and RENO~\cite{R2012} Collaborations, has been interpreted as an effect of neutrino mixing and neutrino mass~\cite{Pontecorvo,GiKi}. Among other things, this event set the quite relevant question of whether the neutrinos correspond to Dirac or Majorana fermions. Clues to the answer might come from experimental searches of the elusive neutrinoless double beta decay. It has been pointed out that the electromagnetic properties of massive neutrinos are very different depending on whether these fermions are of Dirac or Majorana type~\cite{Kayser,shrockacio,Gi1,Gi2,Gi3}, but they are elusive and difficult to analyze. An important aspect of neutrino mixing is that the measurement of a nonzero value of the $\theta_{13}$ mixing angle rendered it a source of $CP$ violation. The Pontecorvo-Maki-Nakagawa-Sakata mixing matrix~\cite{MNS,Pontecorvo2} is able to introduce violations of such invariance by means of one complex phase, if neutrinos are Dirac fermions, or even three phases, in case that neutrinos are Majorana fermions~\cite{GiKi,Petcov}.
\\

Certainly, searches for deviations from Standard Model predictions deserve much attention. The exploration of processes that are quite suppressed, or even forbidden, in the Standard Model may eventually find hints about some theory describing nature beyond this low-energy description.
According to Sakharov criteria, the  nonconservation of $CP$ invariance is a necessary requirement for the baryon asymmetry to occur~\cite{Sakharov}.
The violation of $CP$ symmetry is indeed an effect that is included in the Standard Model, though in small amounts, by the complex phase of the Cabibbo-Kobayashi-Maskawa matrix~\cite{Cabibbo,KobMas}. Studies aimed at other sources of $CP$ violation, from physics beyond the Standard Model, constitute an active topic nowadays. In particular, diverse investigations as, for instance, those performed in Refs.~\cite{CPedm1,CPedm2,CPedm3,CPedm4,AppPS,CPedm5,CPedm6,CPedm7,CPedm8,CPedm9,CPedm10,CPedm11,CPedm12}, have explored the generation of electric dipole moments of elementary particles through $CP$ violation characterizing new-physics formulations. Among all the electric dipole moments of elementary particles, that of the electron is, doubtless, the one which has been most stringently bounded~\cite{PDG,ACME}.
\\

Violation of $CP$ invariance in neutrino mixing might induce electric dipole moments of charged leptons.
With this motivation, we explore, in the present paper, the impact of general lepton charged currents on the electromagnetic form factors of Standard Model charged leptons at one loop. The charged currents that we consider involve Standard Model charged leptons, $l_\alpha$, a heavy charged gauge boson, $W'$, and a set of heavy Dirac neutrinos, $N_j$.
We find that the resulting contributions to diagonal and transition electric and magnetic moments are free of ultraviolet divergences.
Being aware that, in general, masses originate in spontaneous symmetry breaking, we assume that the masses $m_{W'}$ and $m_j$, of the $W'$ boson and the heavy neutrinos $N_j$, grow like $\propto\Lambda$, with some high-energy scale $\Lambda$. This allows us to ensure that the contributions from any neutrino $N_j$, through general charged currents, to 
electric and magnetic moments, both diagonal and of transition type, that are featured in the vertex $\gamma\,l_\alpha l_\beta$ decouple as $\Lambda\to\infty$. 
\\

Our investigation of the electric dipole moments of charged leptons, in this context of general charged currents, shows that these quantities arise at the one-loop level if three conditions are met: 1) the charged currents violate $CP$; 2) the axial and vector terms in the general currents differ from each other; and 3) the neutrinos are massive. This is in contrast to the Standard Model contributions from the Cabibbo-Kobayashi-Maskawa phase to the electric dipole moment of the electron, which vanish even at the three-loop order~\cite{PoKh} and produce, at four loops, the tiny value $\sim{\cal O}(10^{-44})\,e\cdot{\rm cm}$~\cite{PoRi}. Being the difference among vector and axial charged currents a necessary condition to produce one-loop electric dipole moments, we use the upper limit, of order $10^{-{\rm 29}}\,e\cdot{\rm cm}$,  on the electric dipole moment of the electron~\cite{PDG,ACME} to estimate that such differences cannot be larger than $\sim10^{-10}-\sim10^{-7}$ for TeV-sized heavy-neutrino masses, with the $W'$ mass lying within $0.45\,{\rm TeV}-7\,{\rm TeV}$. 
\\

We enquire into the contributions to the anomalous magnetic moments of Standard Model charged leptons, for which we consider first a scenario featuring a heavy-neutrino mass spectrum that is quasidegenerate. We also explore what happens if two neutrinos have masses that are quasidegenerate, but they are different from the mass of a third neutrino. The difference among axial and vector terms of general charged currents, which is essential for electric dipole moments to exist, produces a subleading contribution to the anomalous magnetic moments. In both situations, the contributions to the anomalous magnetic moments turn out to be small in the case of the electron and the muon. Concerning the anomalous magnetic moment of the tau lepton, the contributions range from $\sim10^{-10}$ to $\sim10^{-8}$, which coincides with values that have been reported for diverse models of new physics. In the case of a quasidegenerate set of neutrino masses, we find that the contributions to the anomalous magnetic moments of all the charged leptons share the same sign. Contrastingly, in the second scenario the sign of the contributions can be different, which is entirely determined by the specific texture of neutrino mixing.
\\

We calculate and analyze the flavor-changing decay $\mu\to e\gamma$, which is forbidden in the Standard Model, but whose presence is allowed by general charged currents in which neutrinos mix.
We find a branching ratio that is given in terms of transition magnetic and electric moments.
As with the anomalous magnetic moments, the differences between axial and vector terms of the general charged currents do not produce, in most scenarios, the dominant effects. 
We show that a quasidegenerate spectrum of neutrino masses renders the impact of such differences dominant, thus suppressing the contributions to the transition moments.
In a scenario of two quasidegenerate neutrino masses and one nondegenerate mass, where such suppression does not happen, we have verified that certain regions of neutrino masses and $m_{W'}$ keep the branching ratio below the current upper bound on this flavor changing process, which is of order $10^{-13}$~\cite{PDG}. We show that these regions are wider for larger values of the $W'$ boson mass.
\\

The paper has been organized in the following manner: in Section~\ref{1loopcalc}, we define our framework and sketch the calculation of the one-loop contributions to the flavor changing electromagnetic vertex $\gamma\,l_\alpha l_\beta$; Section~\ref{1loopedms} is dedicated to the one-loop electric dipole moments contributions, where the upper bound on the difference among vector and axial currents is derived for the case of the electron; in Section~\ref{1loopmms}, we explore the contributions to magnetic moments, which we estimate for all the Standard Model charged leptons for two scenarios of neutrino masses; the decay $\mu\to e\gamma$ is calculated, analyzed and discussed in Section~\ref{flvdecay};  and finally, Section~\ref{conclusions} is used to present our conclusions.

\section{Electromagnetic moments from general charged currents}
\label{1loopcalc}
We start by considering the general set of charged currents (CC)
\begin{equation}
{\cal L}_{\rm CC}=\frac{1}{2\sqrt{2}}\sum_j\sum_\alpha\Big[ W'^+_\rho\,\bar{N}_j\,\gamma^\rho(v_{j\alpha}-a_{j\alpha}\gamma^5)l_\alpha+{\rm H.\,c.} \Big],
\label{genCCs}
\end{equation}
where $\alpha=e,\,\mu,\,\tau$ is a flavor index, so that $l_\alpha$ represents charged leptons, and $j=1,2,3$ runs over heavy Dirac neutrinos $N_j$. We are assuming that $v_{j\alpha}\ne \pm a_{j\alpha}$, which occurs, for instance, if the $W'$ originates from a mixing of charged gauge bosons.
The set of coefficients $v_{j\alpha}$ and $a_{j\alpha}$ implicitly bear all the information about heavy-neutrino mixing, which we assume to violate $CP$ invariance. While we are restricting our study to heavy neutrinos, note that, in a more general context, the sum over $j$ could involve both heavy and light neutrinos.
In such case, the set of coefficients $v_{j\alpha}$ and $a_{j\alpha}$, in Eq.~(\ref{genCCs}), could be viewed as entries of nonsquare complex matrices, which thus would not be restricted to be unitary, as is the case of the neutrino mass model analyzed in Ref.~\cite{pilaftsis}. The Greek index $\rho$, in the $W'$ charged boson field, labels spacetime coordinates, with a sum over any pair of repeated indices. A more general set of charged currents could include, in addition, other gauge bosons, as it is the case of the charged currents that were considered in Ref.~\cite{DeNo} to calculate Majorana neutrino magnetic moments in {\it left-right models}~\cite{MoSe}.
\\

\begin{figure}[!ht].
\center
\includegraphics[width=4cm]{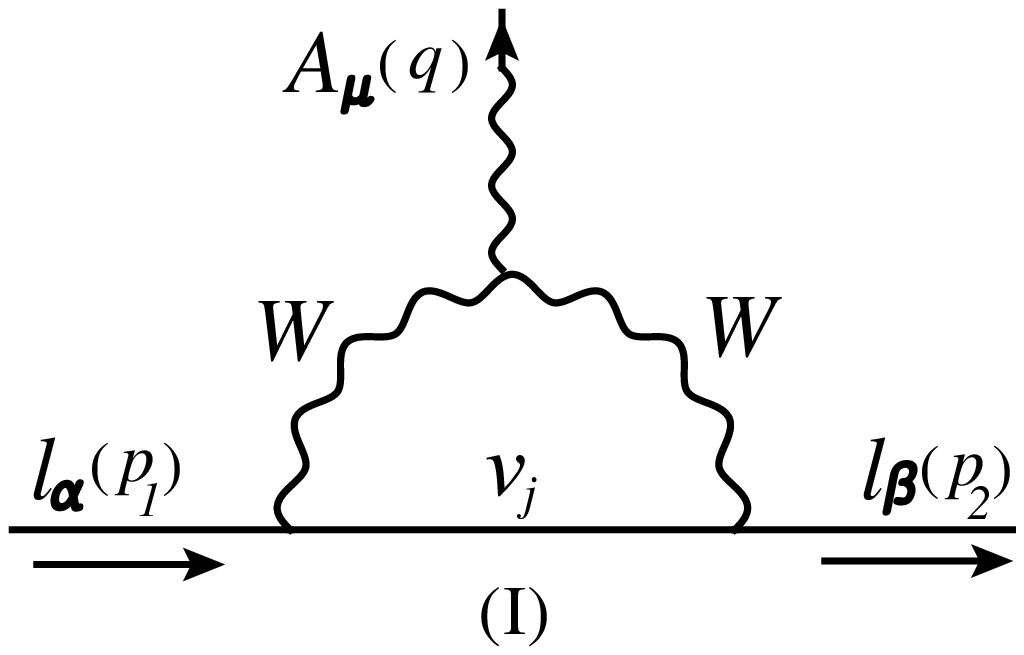}
\\
\vspace{0.4cm}
\includegraphics[width=3.6cm]{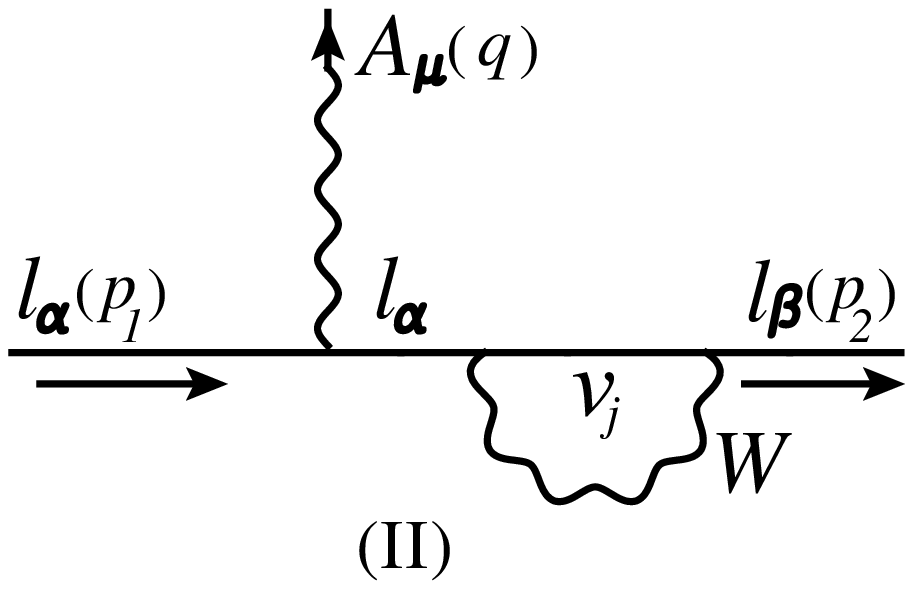}
\hspace{0.4cm}
\includegraphics[width=3.6cm]{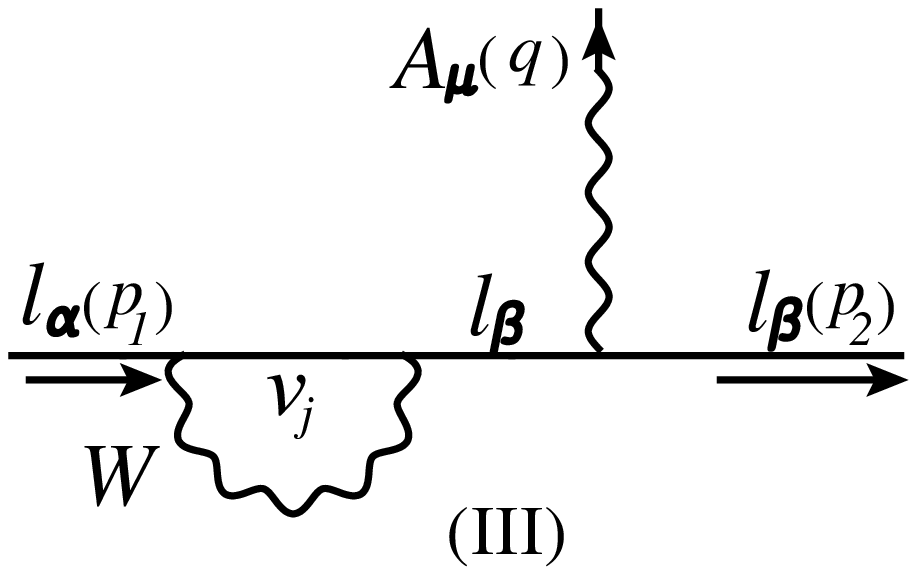}
\caption{\label{emffdiagrams} One-loop diagrams contributing to the flavor-changing electromagnetic vertex.}
\end{figure}

The one-loop contributions from the charged currents given in Eq.~(\ref{genCCs}) to the electromagnetic vertex $\gamma\,l_\alpha l_\beta$, with $l_\alpha$ and $l_\beta$ being either equal or different, emerge from the Feynman diagrams shown in Fig.~\ref{emffdiagrams}.
We perform this calculation in the unitary gauge. Taking the external particles on shell, we find that the resulting vertex function has the well-known structure of the electromagnetic vertex parametrization, which is, for instance, provided in Refs.~\cite{CPedm2,NPR} and which reads\footnote{As it is shown in Refs.~\cite{MNTT1,MNTT2}, violation of Lorentz invariance allows a richer structure of this parametrization.}
\begin{eqnarray}
\Gamma^{\alpha\beta}_\mu&=&ie\Big[ \gamma_\mu(f^{\rm V}_{\alpha\beta}-f^{\rm A}_{\alpha\beta}\,\gamma_5 )
\nonumber \\ &&
-\sigma_{\mu\nu}q^\nu\left( i\frac{\mu_{\alpha\beta}}{m_\alpha+m_\beta}-\frac{d_{\alpha\beta}}{e}\,\gamma_5 \right) \Big].
\label{emvrtxsctr}
\end{eqnarray}
Here, $e$ denotes the unit electric charge (positive), and $m_\alpha$ and $m_\beta$ are the masses of external charged leptons $l_\alpha$ and $l_\beta$. The parameters $f_{\alpha\beta}^{\rm V}$, $f_{\alpha\beta}^{\rm A}$, $\mu_{\alpha\beta}$, and $d_{\alpha\beta}$ are, respectively, the one-loop contributions to electric charge, axial current, anomalous magnetic, and electric dipole moments.
Even though this calculation was carried out on shell and in a specific gauge, all the factors in Eq.~(\ref{emvrtxsctr}) are complicated functions of the masses of all the fields involved in the contributing diagrams (see Fig.~\ref{emffdiagrams}). These are the $W'$ boson mass $m_{W'}$, the masses $m_\alpha$ and $m_\beta$ of charged leptons, and the $N_j$ neutrino masses $m_j$.
\\

We have verified that for $\alpha\ne\beta$ the contributions to $f_{\alpha\beta}^{\rm V}$ and $f_{\alpha\beta}^{\rm A}$ vanish exactly. If $\alpha=\beta$ the corresponding nonzero contributions contain ultraviolet divergences, but they are expected to be absorbed by renormalization. The factors $\mu_{\alpha\beta}$ and $d_{\alpha\beta}$ corresponding to $\alpha\ne\beta$ are respectively called {\it transition magnetic moments} and {\it transition electric moments}. On the other hand, the $\mu_{\alpha}\equiv\mu_{\alpha\alpha}$ are the {\it anomalous magnetic moments} and the $d_\alpha\equiv d_{\alpha\alpha}$ are the {\it electric dipole moments}.
To calculate these loop contributions we utilized the Passarino-Veltman method~\cite{PaVe}, which means that all these electromagnetic moments are expressed in terms of two-point and three-point scalar functions, $B_0$ and $C_0$. While the $C_0$ functions are ultraviolet finite, the $B_0$ functions include such type of divergences. In the dimensional regularization approach~\cite{BoGi} any $B_0$ function can be expressed as~\cite{tHV} $B_0=\Delta_{\rm div}+f_{\rm fin}$, with all the ultraviolet divergences and the logarithmic cutoff dependence contained in $\Delta_{\rm div}$, which is shared by all the two-point functions. Taking advantage of this generic form of the $B_0$ functions, we have checked that all ultraviolet divergences in both transition and diagonal magnetic and electric moments are exactly eliminated, thus yielding finite results for these quantities. This is consistent because the magnetic and electric dipole Lorentz structures are exclusively generated at the loop level.
\\

After performing the calculation, we found it convenient to write down the magnetic and electric moments as
\begin{eqnarray}
\mu_\alpha&=&\sum_j\left(\, |a_{j\alpha}|^2M^{\rm A}_{j\alpha}+|v_{j\alpha}|^2M^{\rm V}_{j\alpha} \,\right),
\label{amm}
\\ \nonumber \\
d_\alpha&=&\sum_ji\big(v_{j\alpha}a^*_{j\alpha}-a_{j\alpha}v_{j\alpha}^* \big)\,D_{j\alpha}\cdot e\,,
\label{edm}
\\ \nonumber \\
\mu_{\alpha\beta}&=&\sum_j\big[ \,a_{j\alpha}\,a^*_{j\beta}\,M^{\rm A}_{j,\alpha\beta}+v_{j\alpha}\,v^*_{j\beta}\,M^{\rm V}_{j,\alpha\beta}\, \big],
\label{tmdm}
\\ \nonumber \\
d_{\alpha\beta}&=&\sum_j i\big[ a_{j\alpha}\,v^*_{j\beta}\,D_{j,\alpha\beta}+v_{j\alpha}\,a^*_{j\beta}\,\overline{D}_{j,\alpha\beta} \big]\cdot e,
\label{tedm}
\end{eqnarray}
each of them with a sum over the index $j$, because of neutrinos circulating in the loops. All the dependence of these electromagnetic moments on Passarino-Veltman functions and, in general, on the masses of particles (see Fig.~\ref{emffdiagrams}), lies within the real-valued factors $M^{\rm A}_{j\alpha}$, $M^{\rm V}_{j\alpha}$, $D_{j\alpha}$, $M^{\rm A}_{j,\alpha\beta}$, $M^{\rm V}_{j,\alpha\beta}$, $D_{j,\alpha\beta}$, and $\overline{D}_{j,\alpha\beta}$. The explicit expressions of all these factors, in terms of Passarino-Veltman scalar functions, are provided in Appendix~\ref{anex}.
Eq.~(\ref{amm}) shows that the anomalous magnetic moments $\mu_\alpha$ are real quantities, which also occurs with the electric dipole moments $d_\alpha$, Eq.~(\ref{edm}), whose terms within the neutrino sum are proportional to $\mathfrak{Im}(v_{j\alpha}^*a_{j\alpha})$.
The transition moments $\mu_{\alpha\beta}$ and $d_{\alpha\beta}$ are, in general, complex numbers.
Let us point out that the charged currents provided in Eq.~(\ref{genCCs}) violate $CP$ invariance if at least $v_{j\alpha}$ or $a_{j\alpha}$ is a complex number, but they preserve this symmetry if both of them are real quantities~\cite{DeNo}. It can be appreciated from Eq.~(\ref{edm}) that the contributions to the electric dipole moment of a charged lepton $l_\alpha$ are nonzero only if $CP$ is violated by the charged currents of Eq.~(\ref{genCCs}) and $v_{j\alpha}\ne\pm a_{j\alpha}$.
Eqs.~(\ref{amm}), (\ref{tmdm}), and (\ref{tedm}) clearly show that conservation of the $CP$ symmetry and/or fulfillment of the condition $v_{j\alpha}=\pm a_{j\alpha}$ does not forbid the existence of contributions to all other electromagnetic moments. In particular, the status of $CP$ symmetry, in this context, is completely irrelevant to the anomalous magnetic moment $\mu_\alpha$, as it can be observed in Eq.~(\ref{amm}).

\section{Electric dipole moments at one loop}
\label{1loopedms}
In this section we turn our attention to the diagonal electric dipole moments, whose general structure is the one given in Eq.~(\ref{edm}). Since the assumption that $v_{j\alpha}\ne a_{j\alpha}$ is a necessary requirement to have nonzero electric dipole moments, we find it suitable to introduce the difference $\Delta_{j\alpha}\equiv a_{j\alpha}-v_{j\alpha}$. Furthermore, because $CP$ violation is also necessary, we assume that $a_{j\alpha}$ and $v_{j\alpha}$ are complex quantities. The diagonal electric moments that we just showed, in Eq.~(\ref{edm}), are sums of contributions from the massive neutrinos $N_j$. For each of such contributions we use the notation $d_{j\alpha}$, so that the total contribution is expressed as $d_\alpha=\sum_jd_{j\alpha}$. 
\\

In Ref.~\cite{BGLMT}, the Higgs decay into two quarks, in the context of the Standard Model, was recently revisited. An interesting element of this study was a ``heavy mass limit'', in which the masses of internal quarks, $m_{q_{\rm int}}$, and the $W$ boson mass, $m_W$, were taken to be the same for a very large electroweak scale $v$, that is, $m_W=m_{q_{\rm int}}$ for very large $v$. It was then observed and discussed that, as long as this condition is fulfilled, the decay amplitude $H\to q_iq_j$ goes to zero if $v\to\infty$, which was used as a consistency check. In the present paper, we assume that some high-energy scale, $\Lambda$, is associated with the generation of the masses $m_j$ and $m_{W'}$, by spontaneous symmetry breaking, so that $m_{W'}$ and $m_j$ grow with the same scale $\Lambda$. In this context, we reasonably assume that for large $\Lambda$ the relation $m_{W'}\approx\kappa_j m_j$ holds, with $\kappa_j$ independent of $\Lambda$. This allows us to write any contribution $d_{j\alpha}$ as
\begin{eqnarray}
d_{j\alpha}&\approx&\frac{2e|\Delta_{j\alpha}||v_{j\alpha}|\sin\phi_{j\alpha}}{(16\pi)^2\kappa_j^2(\kappa_j^2-1)^4m_j}
\Big\{
(\kappa_j-1)(4\kappa_j^6-15\kappa_j^4
\nonumber \\ &&
+12\kappa_j^2
+6\kappa_j^2\log\kappa_j^2-1)+\frac{1}{2}\frac{m_\alpha^2}{m_j^2}\Big[ 2\kappa_j^6+19\kappa_j^4
\nonumber\\ &&
-14\kappa_j^2-7-2(6\kappa_j^4+8\kappa_j^2+1)\log\kappa_j^2 \Big]
\Big\},
\label{dja}
\end{eqnarray}
for small $m_\alpha$. From this expression of $d_{j\alpha}$, it is clear that the contributions to the electric dipole moment $d_\alpha$ decouple, since $d_{j\alpha}\to0$ for $m_j\to\infty$. Note that the $\kappa_j$ factors are independent of the flavor of the charged lepton. The angle $\phi_{j\alpha}$, which is part of Eq.~(\ref{dja}), is a phase difference of the complex phases of $v_{j\alpha}$ and $\Delta_{j\alpha}$. This phase difference is one of the elements that determines the sign and the magnitude of any contribution to the electric dipole moment $d_\alpha$. In particular, if for some $j$ the coefficients $a_{j\alpha}$ and $v_{j\alpha}$ were $CP$ preserving, then $\phi_{j\alpha}=0$, which, consistently, would eliminate the corresponding $d_{j\alpha}$ contribution. 
\\

So far, experiments have not observed electric dipole moments of elementary particles, and this lack of measurements has been translated into upper bounds.
The electric dipole moment of the electron has received special attention among all electric dipoles of elementary particles, being the one which has been most stringently bounded.
Experiments with thallium atoms and ytterbium fluoride molecules achieved upper bounds of order $10^{-27}e\cdot{\rm cm}$ on $|d_e|$~\cite{deth,deytt1,deytt2}.
The current champion, however, is the upper bound recently established by the ACME Collaboration~\cite{ACME}, which reached an important improvement finding $|d_e|<8.7\times10^{-29}e\cdot {\rm cm}$, at 90\% C.L. The huge difference, of 15 orders of magnitude, between the Standard Model contribution~\cite{PoRi} and the current experimental sensitivity has motivated the introduction of new $CP$-violating physics, pursuing less suppressed values for this observable. Particularly, the authors of Ref.~\cite{NGNG} asserted that in the presence of Majorana neutrinos, a two-loop contribution to the electric dipole moment of the electron is produced. Other investigations concerning electric dipole moments and massive neutrinos were carried out in Refs.~\cite{Valle,EHLR}.
\\

Now we discuss the contribution $d_\alpha$ to the electric dipole moment of the charged lepton $l_\alpha$ by exploring a scenario in which the masses $m_j$, of the neutrinos $N_j$, constitute a quasidegenerate spectrum, that is, $m_j\approx m_k$, for any $j$ and $k$. With this in mind, we consider some mass $m_N$ such that $m_N\approx m_j$ for any $j$. Neglecting terms of order $m_\alpha^2/m_j^2$, which are suppressed subleading contributions, and using the aforementioned upper bound on $|d_e|$, by the ACME Collaboration, we set, in this context of quasidegenerate heavy-neutrino masses, the inequality
\begin{widetext}
\begin{equation}
\left|\sum_j|\Delta_{je}||v_{je}|\sin\phi_{je}\right|\lesssim5.56\times10^{-12}\frac{m_{W'}^2}{m_N}\left| \frac{(m_{W'}^2-m_N^2)^3}{m_N^6-12m_N^4m_{W'}^2+15m_N^2m_{W'}^4-4m_{W'}^6+6m_N^4m_{W'}^2\log\left( \frac{m_N^2}{m_{W'}^2} \right)} \right|{\rm GeV}^{-1}.
\label{edmbounder}
\end{equation}
\end{widetext}
Pertaining to this inequality, in Table~\ref{edmboundtab} we provide, for illustrative purposes, some values of its right-hand side for some choices of the masses $m_{W'}$ and $m_N$. There, the first column has values of the $m_{W'}$ mass, whereas all the other cells represent the values of the upper bounds on the factor $|\sum_j|\Delta_{je}||v_{je}|\sin{\phi_{je}}|$ for each pair ($\,m_N,\,m_{W'}$) that is represented by a column $m_N$ and a row corresponding to a value of $m_{W'}$.
According to this table, the right-hand side of Eq.~(\ref{edmbounder}) lies within $\sim10^{-10}$ to $10^{-7}$, for $m_N$ between $0.5\,{\rm TeV}$ and $6\,{\rm TeV}$, and $m_{W'}$ between $0.45\,{\rm TeV}$ and $6\,{\rm TeV}$. The values of the phase differences $\phi_{j\alpha}$ may yield either constructive or destructive effects in the left-hand side of this equation, which would, respectively, increase or reduce its size. Moreover, the factors $v_{j\alpha}$ also play a role in the definition of the size of the terms that contribute to this factor. Nevertheless, in most scenarios this equation should be understood as a guide telling us how small the $|\Delta_{je}|$ factors are, or, in other words, how close the $a_{je}$ and $v_{je}$ are to each other. We could say that, except for specific scenarios in which for some $j$ we have $|v_{je}|\approx0$ or $\sin\phi_{j\alpha}\approx0$, or when a set of particular values of the phases $\phi_{j\alpha}$ yields a destructive effect, the upper bound on each factor $|\Delta_{je}|$ is quite similar to those shown in Table~\ref{edmboundtab}.
\begin{table}[!ht]
\centering
\begin{tabular}{|c|c|c|c|}
\hline
$m_{W'}$ & $m_N=0.5$ TeV & $m_N=2$ TeV & $m_N=6$ TeV
\\ \hline
0.45 TeV & $7.78\times10^{-10}$ & $3.60\times10^{-10}$ & $1.67\times10^{-10}$
\\ \hline
3 TeV & $2.55\times10^{-8}$ & $7.44\times10^{-9}$ & $3.69\times10^{-9}$
\\ \hline
7 TeV & $1.37\times10^{-7}$ & $3.57\times10^{-8}$ & $1.44\times10^{-8}$
\\ \hline
\end{tabular}
\caption{\label{edmboundtab} Upper bounds on $|\sum_j|\Delta_{je}||v_{je}|\sin{\phi_{je}}|$ for different values of the masses $m_{W'}$ and $m_N$.}
\end{table}
For a more or less democratic mixing, defining the $v_{je}$ coefficients, we expect a similar conclusion in scenarios that are characterized by nondegenerate sets of heavy-neutrino masses. We also want to point out that, as Eq.~(\ref{edmbounder}) shows, the dominant effects of the contributions $d_{j\alpha}$ are independent of the mass of the charged lepton, which means that the strictness of the bounds on the $|\Delta_{j\alpha}|$ factors is entirely determined by how stringent the bounds are on $|d_\alpha|$. In the review of the Particle Data Group~\cite{PDG}, the interval $d_\mu=(-0.1\pm0.9)\times10^{-19}e\cdot{\rm cm}$, for the electric dipole moment of the muon, has been reported. Since this bound is weaker than that on the electric dipole moment of the electron by $\sim10$ orders of magnitude, Eq.~(\ref{edmbounder}) establishes bounds on the factors $|\Delta_{j\mu}|$ that are much less restrictive, by the same amount, than those for the factors $|\Delta_{je}|$.

\section{Anomalous magnetic moments}
\label{1loopmms}
\begin{figure}[!ht]
\centering
\includegraphics[width=8.5cm]{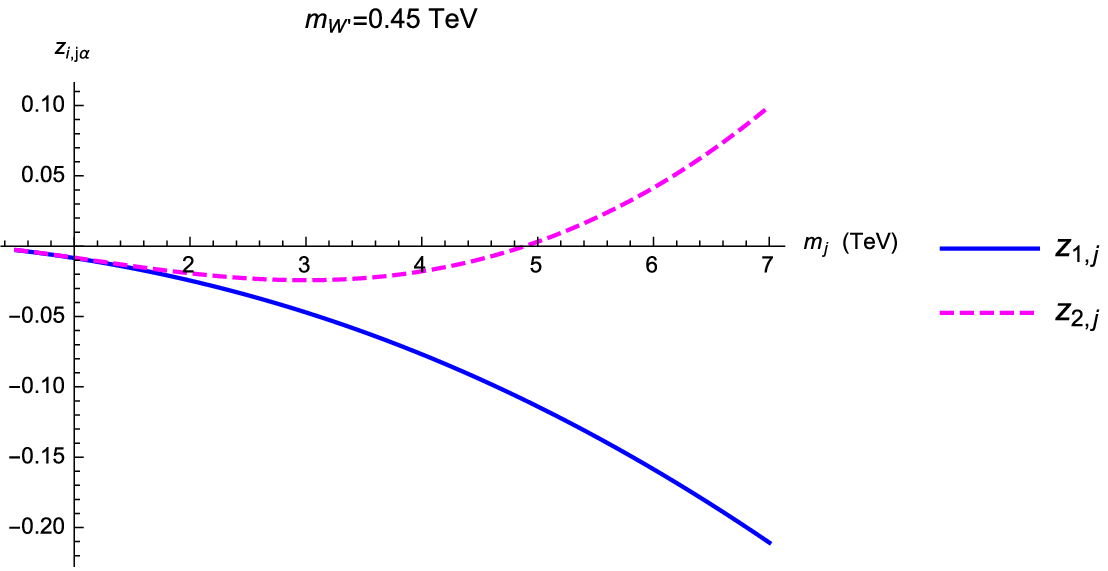}
\\
\vspace{0.5cm}
\includegraphics[width=8.5cm]{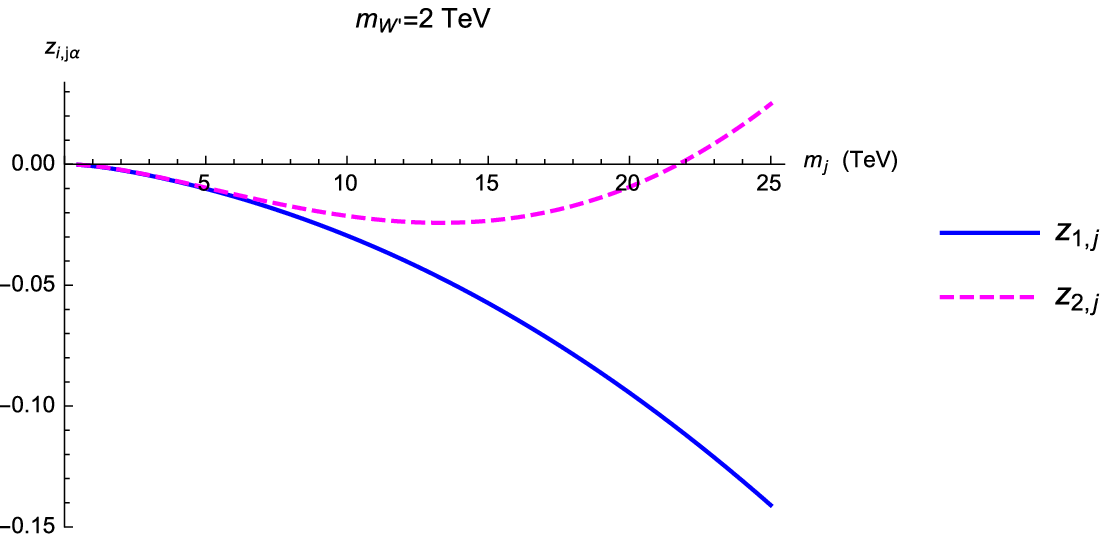}
\caption{\label{z1vz2graphs} Behavior of $z_{1,j}$ and $z_{2,j}$ for mixed $m_{W'}$ and as function of $m_j$. The units used in the $m_j$ axis are TeVs.}
\end{figure}

Analogously to what we did with the electric dipole moments, we denote each neutrino contribution to the magnetic moment $\mu_\alpha$ by $\mu_{j\alpha}$, in which case we have that $\mu_\alpha=\sum_j\mu_{j\alpha}$. It turns out that, for large $\Lambda$, any contribution $\mu_{j\alpha}$ is given by
\begin{equation}
\mu_{j\alpha}\approx2|v_{j\alpha}|\frac{m_\alpha}{m_j}\Big\{ \theta_{j\alpha}|\Delta_{j\alpha}|z_{1,j}+|v_{j\alpha}|\frac{m_\alpha}{m_j}z_{2,j} \Big\},
\label{mja}
\end{equation}
with the factors $z_{1,j}$ and $z_{2,j}$ given by
\begin{eqnarray}
z_{1,j}&=&\frac{-2}{(16 \pi)^2 \kappa _j^2
\left(\kappa _j^2-1\right){}^3}
\Big[
4 \kappa _j^6-15 \kappa _j^4+12 \kappa _j^2
\nonumber \\ &&
+6 \kappa _j^2 \log
\kappa _j^2-1
\Big],
\label{z1ja}
\\
z_{2,j}&=&\frac{-1}{4(16 \pi)^2 \kappa
_j^2 \left(\kappa _j^2-1\right){}^4}\Big[ 26 \kappa _j^8-157 \kappa _j^6
\nonumber \\ &&
+193 \kappa _j^4-91 \kappa
_j^2+29
\nonumber \\ &&
+2 \left(8 \kappa _j^6+30 \kappa _j^4-5 \kappa _j^2+3\right)
\log \kappa _j^2 \Big],
\label{z2ja}
\end{eqnarray}
and where $\theta_{j\alpha}={\rm sgn}(\,|a_{j\alpha}|-|v_{j\alpha}|\,)$
By looking at Eqs.~(\ref{z1ja}) and (\ref{z2ja}), 
we conclude that the contribution displayed in Eq.~(\ref{mja}), to the anomalous magnetic moment $\mu_\alpha$, is a decoupling quantity. As one can appreciate from Eq.~(\ref{mja}), a feature of the contributions $\mu_{j\alpha}$ that contrasts with the electric dipole contributions $d_{j\alpha}$, shown in Eq.~(\ref{dja}), is that their dominant effects include powers of the ratio $m_\alpha/m_j$. About which of the two terms that we included in Eq.~(\ref{mja}) dominates the contribution $\mu_{j\alpha}$, let us point out that, for most situations, the coefficients $z_{1,j}$ and $z_{2,j}$ contribute in similar amounts. In Fig.~\ref{z1vz2graphs}, we show the functions $z_{1,j}$ and $z_{2,j}$ for different values of the neutrino mass $m_j$, with $m_{W'}$ fixed.
In both figures, the blue solid curves represent the coefficient $z_{1,j}$, while for the coefficient $z_{2,j}$ we have used the magenta dashed curves. The plots in the upper figure, corresponding to $m_{W'}=0.45\,{\rm  TeV}$, show that the coefficient $z_{1,j}$ remains negative for all the  values of the neutrino mass $m_j$ between $0.5\,{\rm TeV}$ and $7\,{\rm TeV}$, but $z_{2,j}$, on the other hand, changes from negative to positive, vanishing at $m_j\approx4889.21\, {\rm GeV}$. In the lower figure, for which we considered $m_{W'}=2\,{\rm TeV}$, we observe that the $z_{2,j}$ still changes its sign, but the point at which it is equal to zero shifts to $m_j\approx21729.8\,{\rm GeV}$. Within a narrow neighborhood around the value $m_j$ at which $z_{2,j}$ becomes zero, the first term of Eq.~(\ref{mja}) would be the leading contribution to $\mu_{j\alpha}$, over the second term of this expression. For all  other values of $m_j$, we have to compare the factor $|\Delta_{j\alpha}|$ with $|v_{j\alpha}|(m_\alpha/mj)$. From Eq.~(\ref{edmbounder}) and Table~\ref{edmboundtab}, we note that for many values of $m_{W'}$ and $m_j$ it happens that $(m_e/m_N)\gg|\Delta_{je}|$, in the case of a quasidegenerate spectrum of neutrino masses. As we mentioned during the discussion of the last section, the same conclusion is expected in more general scenarios of neutrino masses. We do not have stringent bounds for $|\Delta_{j\mu}|$ nor $|\Delta_{j\tau}|$. Even so, inspired by the upper bounds that we observed for the factors $|\Delta_{je}|$, we assume that $(m_\alpha/m_j)\gg|\Delta_{j\alpha}|$, for any $\alpha$ and for any $j$. Thus, barring a particular mixing in which $|v_{j\alpha}|\approx 0$, we observe that the dominant contribution in Eq.~(\ref{mja}) comes from its second term. If the latter conditions are fulfilled in a scenario in which $z_{2,j}\approx0$, then the resulting contribution $\mu_{j\alpha}$ gets suppressed by small $|\Delta_{j\alpha}|$ factors.
\\

As a concrete situation, we consider, again, a spectrum of masses of neutrinos that is quasidegenerate and, additionally, we assume that conditions under which the dominant contribution to $\mu_{j\alpha}$ comes from the second term of Eq.~(\ref{mja}) are fulfilled. Under such circumstances, the total contribution to the anomalous magnetic moment $\mu_\alpha$ acquires the form
\begin{eqnarray}
\mu_\alpha&\approx&\frac{1}{2(16\pi)^2}\frac{m_\alpha^2}{m_{W'}^2(m_N^2-m_{W'}^2)^4}
\Big\{
-29m_N^8
\nonumber \\ &&
+91m_N^6m_{W'}^2-193m_N^4m_{W'}^4+157m_N^2m_{W'}^6
\nonumber \\ &&
-26m_{W'}^8+2\Big[ 3m_N^8-5m_N^6m_{W'}^2+30m_N^4m_{W'}^4
\nonumber \\ &&
+8m_N^2m_{W'}^6 \Big]\log\left( \frac{m_N^2}{m_{W'}^2} \right)
\Big\}.
\label{madeg}
\end{eqnarray}
Since, according to Eq.~(\ref{madeg}), the three magnetic moments $\mu_e$, $\mu_\mu$ and $\mu_\tau$ differ from each other only by the factor $m_\alpha^2$, all of them share the same sign. Among the anomalous magnetic moments of charged leptons, the most accurate investigations, in the context of the Standard Model, have been performed in the cases of the muon and the electron~\cite{SMamm1,SMamm2,SMamm3}. The current discrepancies among the values measured by experiments~\cite{expamm1,expamm2,expamm3} and the predictions from the Standard Model are $\Delta \mu_\mu=\mu_\mu^{\rm exp}-\mu_\mu^{\rm SM}=249\,(87)\times10^{-11}$ and $\Delta \mu_e=\mu_e^{\rm exp}-\mu_e^{\rm SM}=-1.06\,(0.82)\times10^{-12}$~\cite{SMamm1,SMamm2}, which means that this is a good place to look for suppressed new physics. Interpreting the contributions given by Eq.~(\ref{madeg}) as lying within such differences between theory and experiment, in the scenario of quasidegenerate neutrino-masses spectrum, would not make sense at all on such theory-experiment grounds. The reason is that the difference $\Delta\mu_\mu$ defines an interval filled with positive numbers and $\Delta\mu_e$ is associated with an interval with points that are exclusively negative, but, as we just emphasized, $\mu_\alpha$ has the same sign for any $\alpha=e,\mu,\tau$, so it is impossible to make these images compatible. In Tables~\ref{tableme}, \ref{tablemm} and \ref{tablemt},
\begin{table}[!ht]
\centering
\begin{tabular}{|c|c|c|c|}
\hline
$m_{W'}$ & $m_N=0.5\,{\rm TeV}$ & $m_N=2\,{\rm TeV}$ & $m_N=6\,{\rm TeV}$
\\
\hline
0.45 TeV & $-5.62\times10^{-15}$ & $-2.52\times10^{-15}$ & $6.02\times10^{-16}$
\\
\hline
3 TeV & $-1.51\times10^{-16}$ & $-1.41\times10^{-16}$ & $-1.01\times10^{-16}$
\\
\hline
7 TeV & $-2.76\times10^{-17}$ & $-2.77\times10^{-17}$ & $-2.48\times10^{-17}$
\\
\hline
\end{tabular}
\caption{\label{tableme} Some values of $\mu_e$ for different choices of $m_j$ and $m_{W'}$.}
\end{table}
\begin{table}[!ht]
\centering
\begin{tabular}{|c|c|c|c|}
\hline
$m_{W'}$ & $m_N=0.5\,{\rm TeV}$ & $m_N=2\,{\rm TeV}$ & $m_N=6\,{\rm TeV}$
\\
\hline
0.45 TeV & $-2.42\times10^{-10}$ & $-1.08\times10^{-10}$ & $2.59\times10^{-11}$
\\
\hline
3 TeV & $-6.51\times10^{-12}$ & $-6.08\times10^{-12}$ & $-4.34\times10^{-12}$
\\
\hline
7 TeV & $-1.19\times10^{-12}$ & $-1.19\times10^{-12}$ & $-1.07\times10^{-12}$
\\
\hline
\end{tabular}
\caption{\label{tablemm} Some values of $\mu_\mu$ for different choices of $m_j$ and $m_{W'}$.}
\end{table}
\begin{table}[!ht]
\centering
\begin{tabular}{|c|c|c|c|}
\hline
$m_{W'}$ & $m_N=0.5\,{\rm TeV}$ & $m_N=2\,{\rm TeV}$ & $m_N=6\,{\rm TeV}$
\\
\hline
0.45 TeV & $-6.80\times10^{-8}$ & $-3.04\times10^{-8}$ & $7.28\times10^{-9}$
\\
\hline
3 TeV & $-1.83\times10^{-9}$ & $-1.71\times10^{-9}$ & $-1.22\times10^{-9}$
\\
\hline
7 TeV & $-3.34\times10^{-10}$ & $-3.35\times10^{-10}$ & $-3.00\times10^{-10}$
\\
\hline
\end{tabular}
\caption{\label{tablemt} Some values of $\mu_\tau$ for different choices of $m_j$ and $m_{W'}$.}
\end{table}
we provide some values for the contributions $\mu_\alpha$, respectively for the cases of the electron, the muon and the tau. By examination of Table~\ref{tablemt}, we find values of $\mu_\tau$ that range from $\sim10^{-10}$ -- $10^{-8}$.
The Standard Model contribution to the anomalous magnetic moment of the $\tau$ lepton was accurately calculated in Ref.~\cite{EiPa}, where the value $\mu^{\rm SM}_\tau=117721(5)\times10^{-8}$ was reported. This value is larger, by several orders of magnitude, than the prediction of the present work, which, however, is not so different from contributions produced by Standard Model extensions. In the case of the Minimal Supersymmetric Standard Model with a mirror fourth generation, the authors of Ref.~\cite{IbrNa} calculated contributions to $\mu_\tau$ that lie between $\sim10^{-9}$ and $\sim10^{-6}$. Anomalous tau magnetic moments arising within seesaw models were explored in Ref.~\cite{Biggio}, where values of orders $10^{-8}$ and $10^{-9}$ were respectively determined for the type I and type III versions of seesaw models. 
It was found in Ref.~\cite{MoTa} that different scenarios allow a spin-0 unparticle to produce contributions to  $\mu_\tau$ in the wide range of values $\sim10^{-10}-\sim10^{-6}$, though the authors of this reference noticed and emphasized that certain scenarios could generate even larger contributions than that of the Standard Model. Ref.~\cite{BMT} includes a calculation of the contributions from scalar leptoquark interactions to the tau anomalous magnetic moment, with values between $\sim10^{-9}$ and $\sim10^{-8}$.
\\

\begin{figure}[!ht]
\centering
\includegraphics[width=8cm]{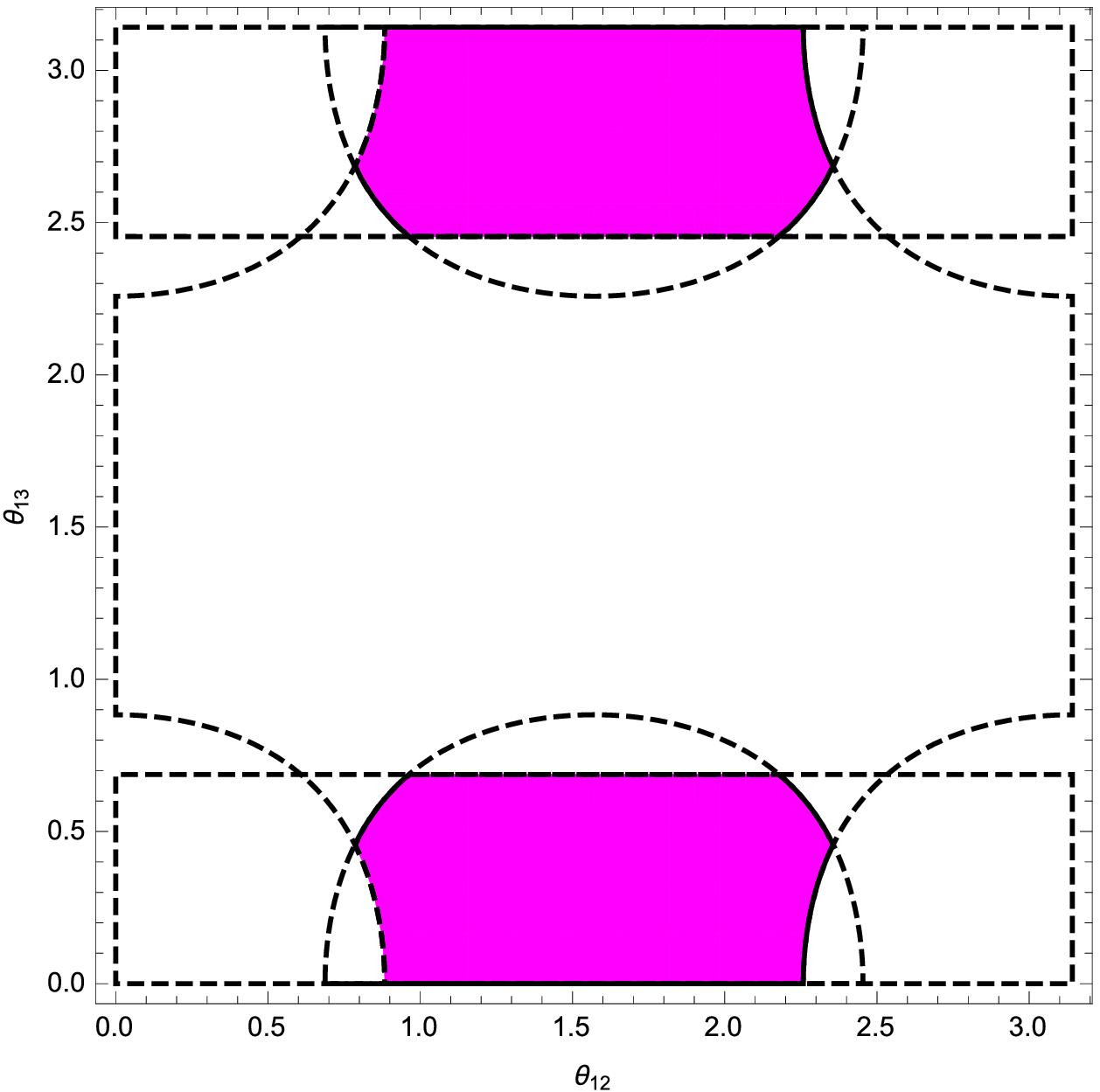}
\\
\vspace{0.4cm}
\includegraphics[width=2.8cm]{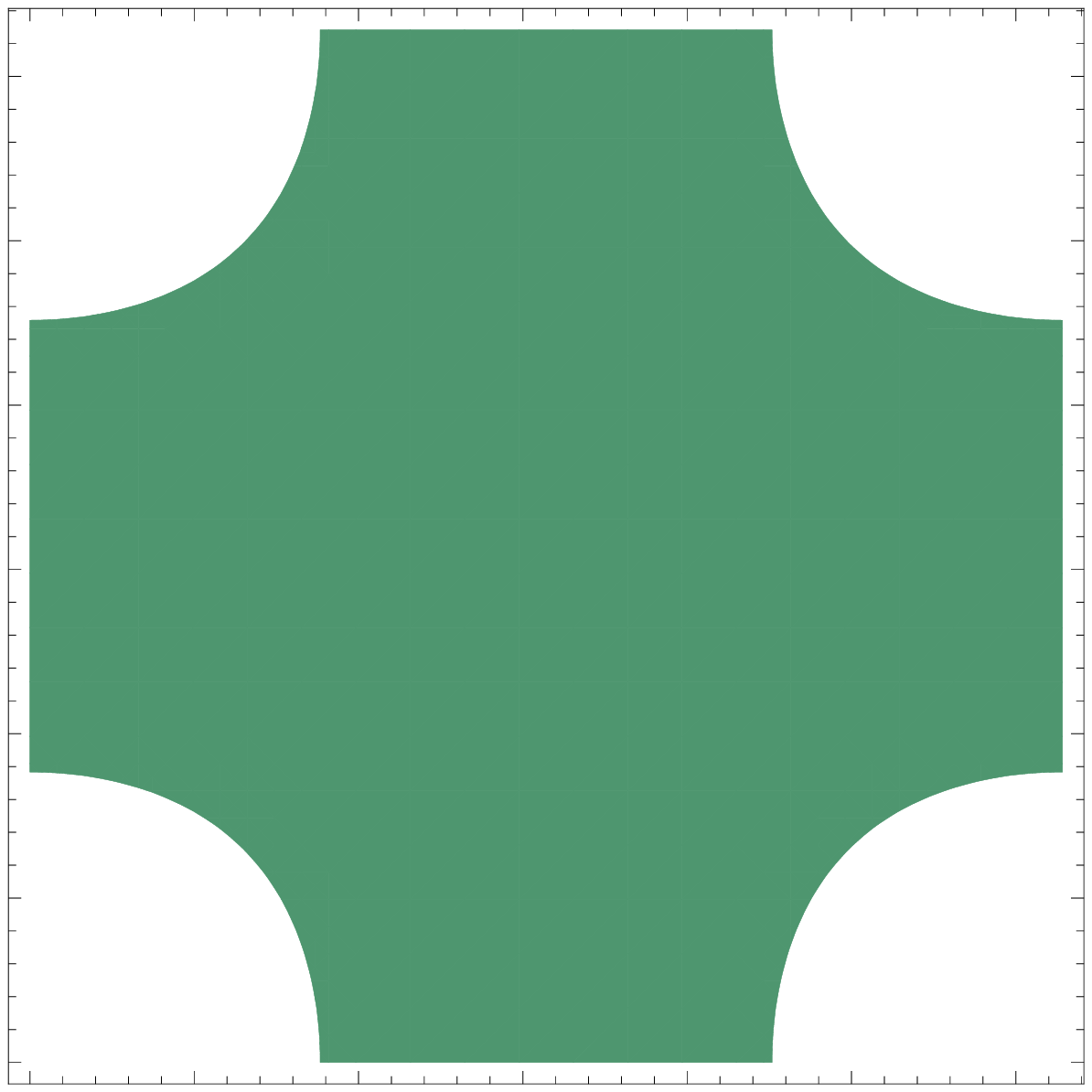}
\includegraphics[width=2.8cm]{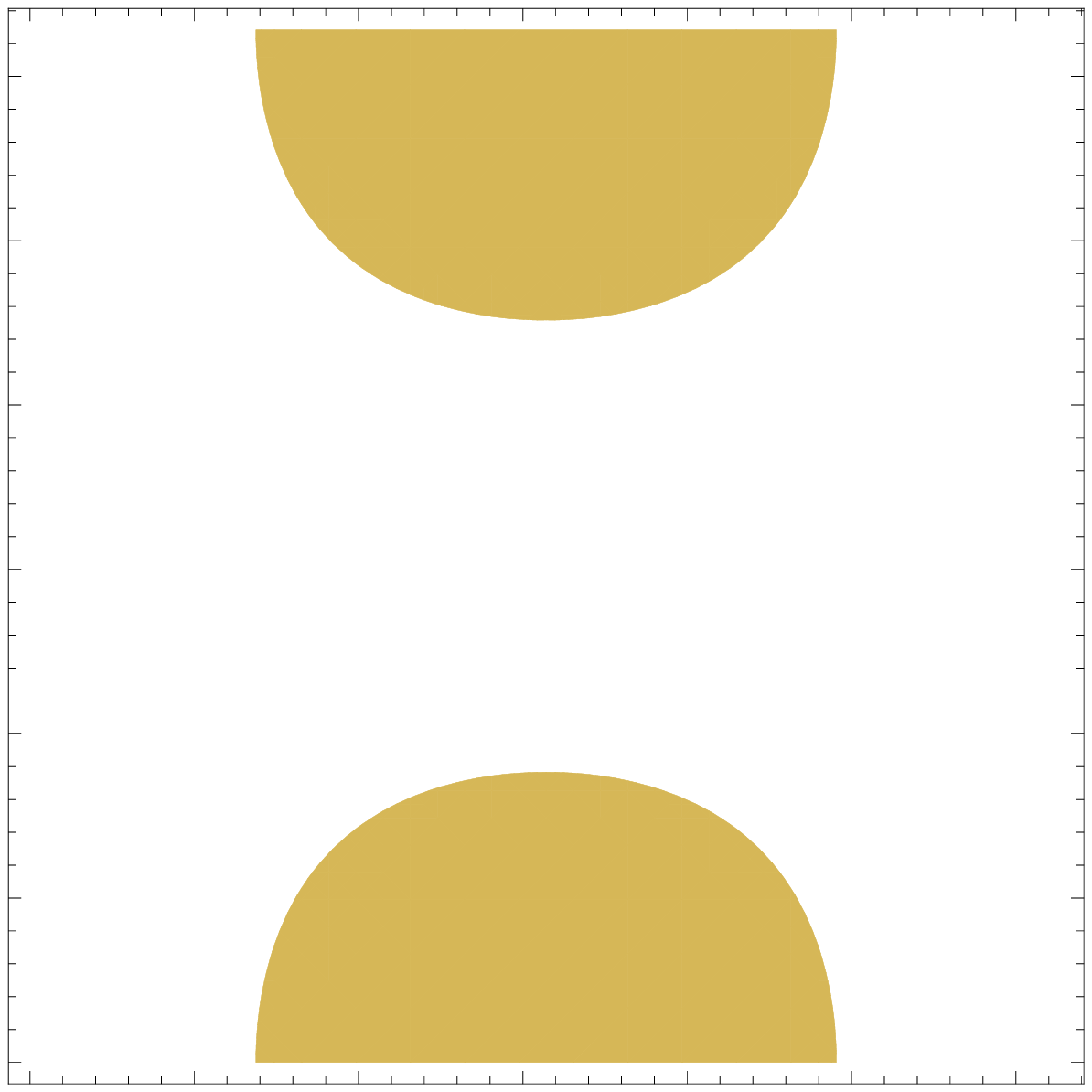}
\includegraphics[width=2.8cm]{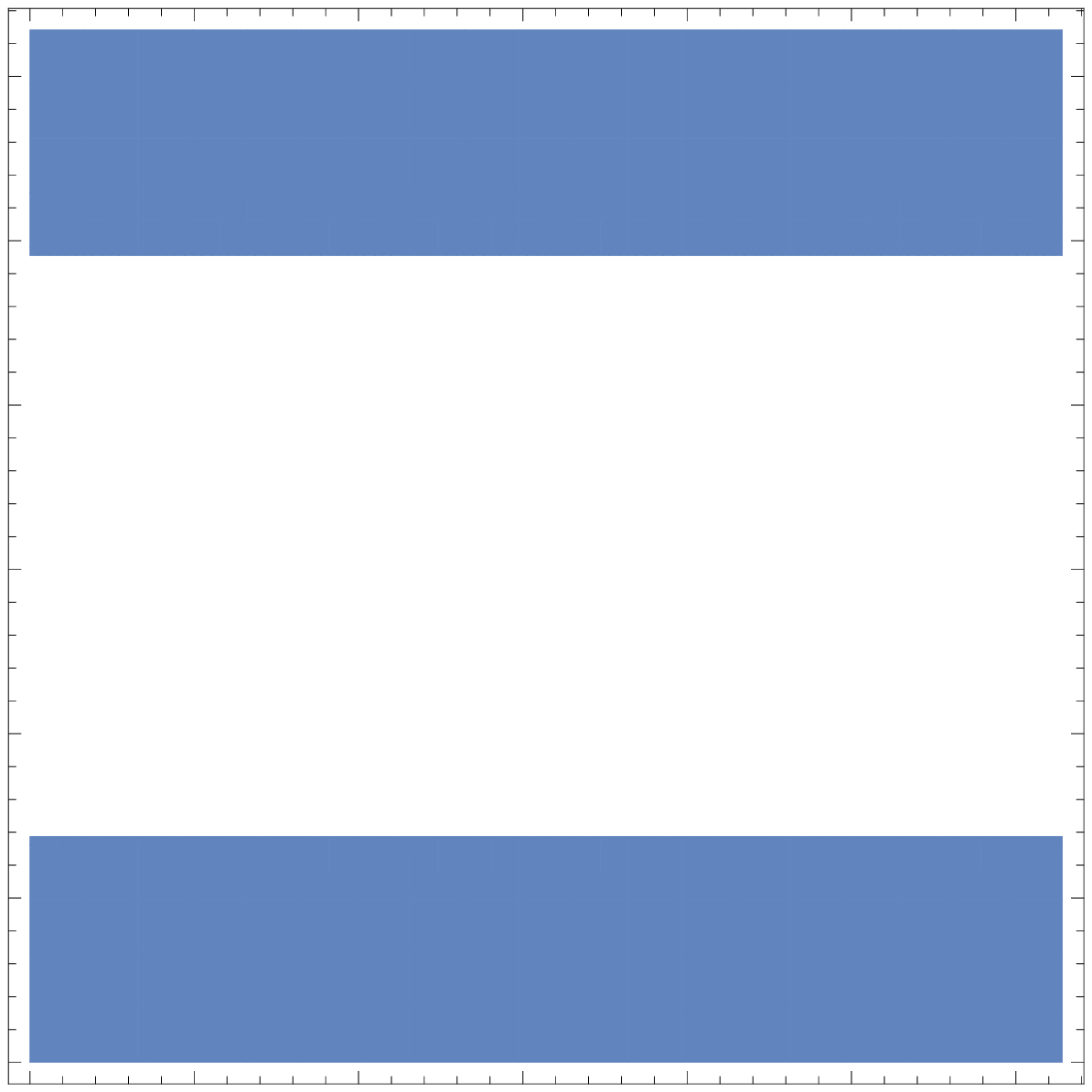}
\caption{\label{msintregion} Intersection of regions in which $\mu_e<0$, $\mu_\mu>0$, and $\mu_\tau<0$ hold.}
\end{figure}

Is it possible to have factors $\mu_\alpha$ in which the signs of the contributions for two different $\alpha$ are different from each other? Indeed, the answer is: yes. To illustrate this point, we investigate a scenario in which the heavy neutrino $N_1$ has a mass $m_1$ that is different from the masses $m_2$ and $m_3$, of the neutrinos $N_2$ and $N_3$, but $m_2\approx m_3$. Then we think of some mass $m_N$ such that $m_N\approx m_2$ and $m_N\approx m_3$. Assuming that the factors $v_{j\alpha}$ are components of a matrix that is, at least, approximately unitary~\footnote{Note that the quantities $v_{j\alpha}$ might involve, for instance, factors from a mixing of charged gauge bosons.} and using the standard parametrization of unitary matrices, in terms of mixing angles $\theta_{ki}$ and one complex phase, we find that
\begin{eqnarray}
\mu_e&\approx&2m_e^2\Big\{\frac{c_{12}^2c_{13}^2}{m_1^2}z_{2,1}+\frac{(1-c_{12}^2c_{13}^2)}{m_N^2}z_{2,N}\Big\},
\\
\mu_\mu&\approx&2m_\mu^2\Big\{ \frac{s_{12}^2c_{13}^2}{m_1^2}z_{2,1}+\frac{(1-s_{12}^2c_{13}^2)}{m_N^2}z_{2,N} \Big\},
\\
\mu_\tau&\approx&2m_\tau^2\Big\{ \frac{s_{13}^2}{m_1^2}z_{2,1}+\frac{c^2_{13}}{m_N^2}z_{2,N} \Big\},
\end{eqnarray}
where we have used the conventional notation $c_{ik}=\cos\theta_{ik}$, $s_{ik}=\sin\theta_{ik}$. In regions of the plane $(m_{W'},m_1)$ and $(m_{W'},m_N)$ in which neither $z_{2,1}$ nor $z_{2,N}$ is close to zero, the sign of each $\mu_\alpha$ is determined by the mixing angles. For instance, take a look at Fig.~\ref{msintregion}, which corresponds to the choices $m_{W'}=0.45\,{\rm TeV}$, $m_1=9\,{\rm TeV}$, $m_N=1.5\,{\rm TeV}$. The upper graph shows the plane $(\theta_{12},\theta_{13})$, with these angles running from 0 to $\pi$. Within such a plane, we have colored in magenta a region in which the conditions $\mu_e<0$, $\mu_\mu>0$, $\mu_\tau<0$ are fulfilled simultaneously. The dashed lines are the borders of the regions corresponding to each of such conditions. 
For clarity purposes, we have added the three lower graphs to Fig.~\ref{msintregion}. They show, separately and in color, the regions of the aforementioned conditions: the green one is the region for $\mu_e<0$; yellow corresponds to the region for $\mu_\mu>0$, and blue has been used for the region in which $\mu_\tau<0$. We have checked that the values of the factors $2z_{2,1}(m_\alpha^2/m_1^2)$ and $2z_{2,N}(m_\alpha^2/m_N^2)$ are similar to those shown in Tables~\ref{tableme}, \ref{tablemm}, and \ref{tablemt}, which means that the contributions $\mu_\alpha$ in this scenario of heavy-neutrino masses are also similar to those in the quasidegenrate spectrum scenario. Even though in the present scenario of neutrino masses it is possible to tune the sign of the contributions to the anomalous magnetic moments $\mu_\alpha$, it is still not correct to interpret them as the differences $\Delta\mu_e$ and $\Delta\mu_\mu$, because the contributions produced by the present model are too small and cannot fall in the regions defined by such data, unless the $m_{W'}$ mass is unacceptably small.

\section{The flavor changing decay $\boldsymbol{\mu\to e\gamma}$}
\label{flvdecay}
The decay rate $\Gamma(l_\alpha\to l_\beta\gamma)$ is expressed in terms of the transition electromagnetic moments $d_{\alpha\beta}$ and $\mu_{\alpha\beta}$, given in Eqs.~(\ref{tmdm}) and (\ref{tedm}).
In such equations we wrote these transition moments in terms of the coefficients $M^{\rm A}_{j,\alpha\beta}$, $M^{\rm V}_{j,\alpha\beta}$, $D_{j,\alpha\beta}$, and $\overline{D}_{j,\alpha\beta}$, which can be conveniently expressed, for small masses $m_\alpha$ and $m_\beta$, as
\begin{equation}
M^{\rm A}_{j,\alpha\beta}\approx\frac{m_\alpha+m_\beta}{m_j}\eta^{(1)}_j+\frac{m_\alpha^2+m_\beta^2}{m_j^2}\eta^{(2)}_j
+\frac{m_\alpha m_\beta}{m_j^2}\eta^{(3)}_j,
\label{MAkj}
\end{equation}
\begin{equation}
M^{\rm V}_{j,\alpha\beta}\approx-\frac{m_\alpha+m_\beta}{m_j}\eta^{(1)}_j+\frac{m_\alpha^2+m_\beta^2}{m_j^2}\eta^{(2)}_j+\frac{m_\alpha m_\beta}{m_j^2}\eta^{(3)}_j,
\label{MVkj}
\end{equation}
\begin{eqnarray}
D_{j,\alpha\beta}&\approx&
\frac{1}{m_j}\omega^{(1)}_j+\frac{m_\alpha^2+m_\beta^2}{m_j^2(m_\alpha-m_\beta)}\omega^{(2)}_j
\nonumber \\ &&
+\frac{m_\alpha m_\beta}{m_j^2(m_\alpha-m_\beta)}\omega^{(3)}_j,
\label{Dkj}
\\ \nonumber \\
\overline{D}_{j,\alpha\beta}&\approx&-\frac{1}{m_j}\omega^{(1)}_j+\frac{m_\alpha^2+m_\beta^2}{m_j^2(m_\alpha-m_\beta)}\omega^{(2)}_j
\nonumber \\ &&
+\frac{m_\alpha m_\beta}{m_j^2(m_\alpha-m_\beta)}\omega^{(3)}_j,
\label{Dbarkj}
\end{eqnarray}
where the factors $\eta^{(n)}_{j}$ and $\omega^{(n)}_{j}$, whose explicit expressions can be found in Appendix~\ref{anex1}, depend only on $\kappa_j$. This means that these factors are independent of the energy scale $\Lambda$, in turn implying that the right-hand sides of Eqs.~(\ref{MAkj}) to (\ref{Dbarkj}) go to zero in the limit in which $\Lambda\to\infty$. Thus, the transition moments $\mu_{\alpha\beta}$ and $d_{\alpha\beta}$ decouple. 
\\

Let us write the transition moments $\mu_{\alpha\beta}$ and $d_{\alpha\beta}$ as
\begin{eqnarray}
\mu_{\alpha \beta}&=&
\sum_j\Big\{
v_{j\alpha}v^*_{j\beta}(M^{\rm A}_{j,\alpha \beta}+M^{\rm V}_{j,\alpha \beta})
\nonumber \\ &&
+\Big(
\Delta_{j\alpha}v^*_{j\beta}+v_{j\alpha}\Delta^*_{j\beta}+\Delta_{j\alpha}\Delta^*_{j\beta}
\Big)M^{\rm A}_{j,\alpha\beta}
\label{mabfordec}
\\ \nonumber \\
d_{\alpha \beta}&=&\sum_jie
\Big\{
v_{j\alpha}v^*_{j\beta}(D_{j,\alpha \beta}+\overline{D}_{j,\alpha \beta})
\nonumber \\ &&
+\Delta_{j\alpha}v^*_{j\beta}D_{j,\beta\alpha }
+v_{j\alpha}\Delta^*_{j\beta}\overline{D}_{j,\alpha \beta}
\Big\}.
\label{dabfordec}
\end{eqnarray}
Looking at Eq.~(\ref{mabfordec}), we wish to emphasize that, according to Eqs.~(\ref{MAkj}) and (\ref{MVkj}), all the terms in ($M^{\rm A}_{j,\alpha\beta}+M^{\rm V}_{j,\alpha\beta}$) that are linear in $m_\alpha/m_j$ and $m_\beta/mj$ cancel exactly, so that the leading contributions in this sum are terms of orders $m_\alpha^2/m_j^2$, $m_\beta^2/m_j^2$, and $(m_\alpha m_\beta)/m_j^2$. In contrast, the sole factor $M^{\rm A}_{j,\alpha\beta}$ has lower-order contributions with respect to $m_\alpha/mj$ and $m_\beta/mj$. Nevertheless, all the terms that are proportional to $M^{\rm A}_{j,\alpha\beta}$ in Eq.~(\ref{mabfordec}) also involve factors $|\Delta_{j\alpha}|$. Recalling that, based on our analysis on the electric dipole moment of the electron, we assumed that $m_\alpha/mj\gg|\Delta_{j\alpha}|$, for any $\alpha$, it turns out that the leading contributions to the whole transition moment $\mu_{\alpha\beta}$ come from the sum of terms $v_{j\alpha}v^*_{j\beta}(M^{\rm A}_{j\alpha\beta}+M^{\rm V}_{j,\alpha\beta})$, in Eq.~(\ref{mabfordec}). Similar arguments lead us to conclude that it is the sum of terms $v_{j\alpha}v^*_{j\beta}(D_{j,\alpha\beta}+\overline{D}_{j,\alpha\beta})$, in Eq.~(\ref{dabfordec}), the one that produces the dominant contributions to the transition electric moments $d_{\alpha\beta}$. The neutrino masses are a main aspect that determines the size of $\mu_{\alpha\beta}$ and $d_{\alpha\beta}$. Concerning this point, it is worth noting that for a quasidegenerate spectrum of neutrino masses it happens that $M^{\rm A}_{j,\alpha \beta}\approx M^{\rm A}_{k,\alpha \beta}$ for any $j$ and $k$, and something analogous for  the factors $M^{\rm V}_{j,\alpha\beta}$, $D_{j,\alpha\beta}$, and $\overline{D}_{j,\alpha\beta}$. In such a case, the condition $\sum_jv_{j\alpha}v^*_{j\beta}=\delta_{\alpha\beta}$ largely suppresses the contributions $\sum_jv_{j\alpha}v^*_{j\beta}(M^{\rm A}_{j,\alpha\beta}+M^{\rm V}_{j,\alpha\beta})$ and $\sum_jie\,v_{j\alpha}v^*_{j\beta}(D_{j,\alpha\beta}+\overline{D}_{j,\alpha\beta})$. Since all other contributions are proportional to factors $|\Delta_{j\alpha}|$ and $|\Delta_{j\beta}|$, we observe that both transition moments $\mu_{\alpha\beta}$ and $d_{\alpha\beta}$ then get simultaneously suppressed. 
\\

\begin{figure}[!ht]
\center
\includegraphics[width=8.5cm]{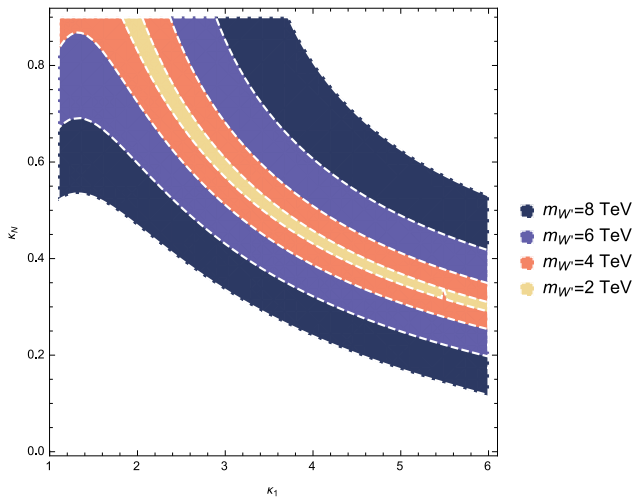}
\caption{\label{regiondecay} Regions, in the $(\kappa_1,\kappa_N)$ plane, in which ${\rm Br}(\mu\to e\gamma)<5.7\times10^{-13}$, barring mixing dependence. The width of each region is determined by the mass $m_{W'}$.}
\end{figure}

Now we concentrate on the case $\alpha=\mu$, $\beta=e$, corresponding to the decay $\mu\to e\gamma$. This process cannot happen in the Standard Model, and even in the minimal extension, where its neutrinos are endowed with masses, the resulting contribution to this decay is tiny, of order $10^{-54}$~\cite{BerCoo}. In order to avoid the suppression of contributions by a quasidegenerate neutrino-mass spectrum, which we described above, we develop the discussion within a context in which the neutrino mass $m_1$ is different from $m_2$ and $m_3$, but $m_2\approx m_3$. As we did in the previous section, we use the mass $m_N$ to characterize $m_2$ and $m_3$. We write the leading contribution to the decay rate $\Gamma(\mu\to e\gamma)$ as
\begin{widetext}
\begin{eqnarray}
\Gamma(l_\mu\to l_e\gamma)&=&\frac{e^2 s^2_{12}c^2_{12}c^4_{13}}{\pi}\frac{(m_\mu^2-m_e^2)^3}{m_\mu^3m_1^4m_N^4}\bigg\{ \bigg[ \frac{(m_\mu^2+m_e^2)(m_N^2\eta^{(2)}_1-m_1^2\eta^{(2)}_N)+m_\mu m_e(m_N^2\eta^{(3)}_1-m_1^2\eta^{(3)}_N)}{m_\mu+m_e} \bigg]^2 
\nonumber \\ &&
+\bigg[
\frac{(m_\mu^2+m_e^2)(m_N^2\omega^{(2)}_1-m_1^2\omega^{(2)}_N)+m_\mu m_e(m_N^2\omega^{(3)}_1-m_1^2\omega^{(3)}_N)}{m_\mu-m_e}
\bigg]^2
\bigg\}.
\label{decayratemuegamma}
\end{eqnarray}
\end{widetext}
The MEG Collaboration has established the most stringent upper bound on the branching ratio for the decay $\mu\to e\gamma$, which they reported to be $5.7\times10^{-13}$~\cite{MEG,PDG}. In Fig.~\ref{regiondecay},
we show the $(\kappa_1,\kappa_N)$ plane, with $\kappa_1$ running from 1.1 to 1.6 and $\kappa_N$ ranging between 0.01 and 0.9. This election of intervals for $\kappa_1$ and $\kappa_N$ describes a set of scenarios in which $m_1<m_{W'}$ and $m_N>m_{W'}$. Within this plane, we have included different colored regions that include all the values $(\kappa_1,\kappa_N)$ for which the branching ratio ${\rm Br}(\mu\to e\gamma)$, calculated with Eq.~(\ref{decayratemuegamma}), remains lower than the aforementioned upper bound, with each of these regions corresponding to different values of the mass $m_{W'}$, of the heavy charged boson $W'$. As we can appreciate from this figure, the larger the mass $m_{W'}$, the wider the region. To devise this graph, we have not considered the angular dependence, on the mixing angles $\theta_{12}$ and $\theta_{13}$ featured in Eq.~(\ref{decayratemuegamma}), so that these regions correspond to optimal values of such mixings. The narrowest region, which we colored in yellow, was plotted for $m_{W'}=2\,{\rm TeV}$. In this region, if a particular value for, say, $\kappa_1$ is fixed, the set of allowed values for $\kappa_N$ lies within a small interval. On the other hand, the widest region ($m_{W'}=8\,{\rm TeV}$), in dark blue, provides more flexibility, since a fixed value $\kappa_1$ imposes minimal restrictions on the allowed values of $\kappa_N$. 
Other physical contexts, in which the values of $(\kappa_1,\kappa_N)$ are different from those shown in Fig.~\ref{regiondecay}, yield analogous regions, in which the same pattern that we just described holds.

\section{Conclusions}
\label{conclusions}
In the light of the confirmation that the phenomenon of neutrino oscillations exists, including neutrino mixing, neutrino mass and $CP$ violation, in this paper we have explored a model in which heavy neutrinos and a heavy $W'$ gauge boson, both originating in some high energy formulation, couple with charged leptons from the Standard Model in a set of general charged currents characterized by vector and axial terms that differ from each other, though by a small amount. We have shown that, as long as 1) such difference is present, 2) $CP$ is violated by these charged currents, and 3) neutrinos are massive, then nonzero contributions from such charged currents to electric dipole moments of  charged leptons arise at one loop.
We have used the most stringent upper bound on the electron electric dipole moment, recently reported by the ACME Collaboration, to estimate that the upper bound on the difference between the vector and axial parts of the charged currents characterizing the electron is within $\sim10^{-10}-10^{-7}$, for heavy neutrino masses in the range $0.5\,{\rm TeV}-7\,{\rm TeV}$ and a mass $m_{W'}$ between $0.45\,{\rm TeV}$ and $7\,{\rm TeV}$.
We have also performed an analysis of the contributions to anomalous magnetic moments in two scenarios with different neutrino mass spectra, which are a quasidegenerate set of heavy-neutrino masses and a spectrum in which two neutrino masses are quasidegenerate and the third one is not close to them. We have determined the size of the contributions from general charged currents to the anomalous magnetic moments of the Standard Model charged leptons. In particular, we provided an estimation of the anomalous magnetic moment of the tau lepton, which turned out to be within the range $\sim10^{-10}-10^{-8}$ for the aforementioned values of the neutrino and $W'$ masses. The last part of the discussion was devoted to flavor-changing decays of Standard Model charged leptons into another charged lepton and a photon. We pointed out that a quasidegenerate spectrum of neutrino masses lowers the value of the contribution to this decay, since in this context the leading contributions are proportional to factors characterizing the difference among axial and vector charged currents, which are tiny. In the case of a more general set of neutrino masses, we showed that in certain regions of neutrino masses, for fixed $W'$ mass, the contributions from the branching ratio ${\rm Br}(\mu\to e\gamma)$ remain below the upper bound reported by the Particle Data Group. We have illustrated that such region widens as we take larger values of the mass $m_{W'}$.
\\


\begin{acknowledgments}
H.N.S. and J.J.T. acknowledge financial support from CONACYT and SNI (M\'exico). H.N.S. also acknowledges financial support from PRODEP (M\'exico), Project DSA/103.5/16/10420.
\end{acknowledgments}

\appendix

\section{Exact analytic expressions of factors in the magnetic and electric moments}
\label{anex}
In Eqs.~(\ref{amm}) to (\ref{tedm}) we gave the general form of the contributions to magnetic and electric moments of charged leptons. These expressions are written in terms of mass-dependent factors $M^{\rm A}_{j\alpha}$, $M^{\rm V}_{j\alpha}$, $D_{j\alpha}$, $M^{\rm A}_{j,\alpha\beta}$, $M^{\rm V}_{j,\alpha\beta}$, $D_{j,\alpha\beta}$, and $\overline{D}_{j,\alpha\beta}$, whose exact analytic expressions are provided in this Appendix. Using the standard definitions 
\begin{widetext}
\begin{eqnarray}
B_0(p^2,m_0^2,m_1^2)&=&\frac{(2\pi\mu)^{4-D}}{i\pi^2}\int d^Dk\,\frac{1}{\big[ k^2-m_0^2 \big]\big[ (k+p)^2-m_1^2 \big]},
\\
C_0(q^2,p^2,(q-p)^2,m_0^2,m_1^2,m_2^2)&=&\frac{(2\pi\mu)^{4-D}}{i\pi^2}\int d^Dk\,\frac{1}{\big[k^2-m_0^2\big]\big[ (k+q)^2-m_1^2 \big]\big[ (k+q-p)^2-m_2^2 \big]},
\end{eqnarray}
\end{widetext}
we have the following expressions:
\begin{widetext}
\begin{eqnarray}
M^{\rm A}_{j\alpha}&=&\frac{1}{{2 (16\pi)^2 m_{\alpha }^2 m_{W'}^2}}
\Big\{
2\left(m_j-m_{W'}\right) \left(m_j+m_{W'}\right) \left(\left(m_j+m_{\alpha}\right){}^2+2 m_{W'}^2\right)B_0(0,m_j^2,m_{W'}^2)+\Big(2m_j^3 m_{\alpha }
\nonumber \\ &&
+m_j^2 \left(3 m_{W'}^2-4 m_{\alpha }^2\right)-2 m_j m_{\alpha }\left(m_{\alpha }^2-5 m_{W'}^2\right)+3 m_j^4+m_{\alpha }^4+15 m_{\alpha }^2m_{W'}^2-6 m_{W'}^4\Big)B_0(0,m_{W'}^2,m_{W'}^2) 
\nonumber \\ &&
-\Big(-2 m_jm_{\alpha }^3+m_{\alpha }^2 \left(13 m_{W'}^2-2 m_j^2\right)+6 m_j m_{\alpha }\left(m_j^2+m_{W'}^2\right)+5 \left(m_j^2 m_{W'}^2+m_j^4-2 m_{W'}^4\right)
\nonumber \\ &&
+m_{\alpha}^4\Big)B_0(m_\alpha^2,m_j^2,m_{W'}^2)
+\left(-m_j-m_{\alpha}+m_{W'}\right) \left(m_j+m_{\alpha }+m_{W'}\right) \Big(4 m_j^3 m_{\alpha }+m_j^2\left(2 m_{\alpha }^2-3 m_{W'}^2\right)
\nonumber \\ &&
+2 m_j m_{\alpha } \left(m_{W'}^2-2 m_{\alpha}^2\right)-3 m_j^4+m_{\alpha }^4-11 m_{\alpha }^2 m_{W'}^2+6 m_{W'}^4\Big)C_0(m_\alpha^2,m_\alpha^2,0,m_{W'}^2,m_j^2,m_{W'}^2)
\nonumber \\ &&
-2m_\alpha^2\Big(\left(m_j+m_{\alpha }\right){}^2+2 m_{W'}^2\Big)
\Big\},
\end{eqnarray}
\begin{eqnarray}
M_{j\alpha}^V&=&
\frac{1}{2(16
\pi)^2 m_\alpha^2m_{W'}^2}\Big\{
2\left(m_j-m_{W'}\right) \left(m_j+m_{W'}\right)
\left(\left(m_j-m_{\alpha }\right){}^2+2 m_{W'}^2\right)B_0(0,m_j^2,m_{W'}^2)+\Big(2 m_j m_{\alpha}^3
\nonumber \\ &&
+m_{\alpha }^2 \left(15 m_{W'}^2-4 m_j^2\right)-2 m_j m_{\alpha
} \left(m_j^2+5 m_{W'}^2\right)+3 \left(m_j^2 m_{W'}^2+m_j^4-2
m_{W'}^4\right)+m_{\alpha }^4\Big)B_0(0,m_{W'}^2,m_{W'}^2)
\nonumber \\ &&
-\Big(2 m_j m_{\alpha }^3+m_{\alpha }^2\left(13 m_{W'}^2-2 m_j^2\right)-6 m_j m_{\alpha }
\left(m_j^2+m_{W'}^2\right)+5 \left(m_j^2 m_{W'}^2+m_j^4-2
m_{W'}^4\right)
\nonumber \\ &&
+m_{\alpha }^4\Big)B_0(m_\alpha^2,m_j,m_{W'}^2)
+\left(m_j-m_{\alpha }+m_{W'}\right)
\left(-m_j+m_{\alpha }+m_{W'}\right) \Big(4 m_j m_{\alpha
}^3+m_{\alpha }^2 \left(2 m_j^2-11 m_{W'}^2\right)
\nonumber \\ &&
-2 m_j m_{\alpha
} \left(2 m_j^2+m_{W'}^2\right)-3 \left(m_j^2 m_{W'}^2+m_j^4-2
m_{W'}^4\right)+m_{\alpha }^4\Big)C_0(m_\alpha^2,m_\alpha^2,0,m_{W'}^2,m_j^2,m_{W'}^2)
\nonumber \\ &&
-2m_\alpha^2\Big(\left(m_j-m_{\alpha }\right){}^2+2 m_{W'}^2\Big)
\Big\},
\end{eqnarray}
\begin{eqnarray}
D_{j\alpha}&=&\frac{m_j}{(16\pi)^2m_\alpha^2m_{W'}^2}
\Big\{
\left(m_j^2-m_{\alpha }^2-4 m_{W'}^2\right)\Big(B_0(m_\alpha^2,m_j^2,m_{W'}^2)-B_0(0,m_{W'}^2,m_{W'}^2)\Big)
\nonumber \\ &&
-\left(-m_j^2 \left(2 m_{\alpha}^2+5 m_{W'}^2\right)+m_j^4+m_{\alpha }^4-3 m_{\alpha }^2m_{W'}^2+4 m_{W'}^4\right)C_0(m_\alpha^2,m_\alpha^2,0,m_{W'}^2,m_j^2,m_{W'}^2)
\Big\},
\end{eqnarray}
\begin{eqnarray}
M^{\rm A}_{j,\alpha\beta}&=&
\frac{1}{(16\pi)^2 m_{\alpha }m_{\beta } (m^2_{\alpha }-m^2_{\beta }) m_{W'}^2}
\Big\{
\left(m^2_{\alpha }-m^2_{\beta }\right) \left(m_{W'}^2-m_j^2\right) \left(\left(m_j+m_{\alpha }\right)\left(m_j+m_{\beta }\right)+2 m_{W'}^2\right)B_0(0,m_j^2,m_{W'}^2)
\nonumber \\ &&
-m_{\beta}\Big(m_j^4 \left(2 m_{\alpha }+m_{\beta}\right)+m_j^3 m_{\beta } \left(m_{\alpha }+m_{\beta }\right)+m_j^2\left(\left(2 m_{\alpha }+m_{\beta }\right) m_{W'}^2-m_{\alpha }\left(m_{\alpha } m_{\beta }+2 m_{\alpha }^2+m_{\beta}^2\right)\right)
\nonumber \\ &&
-m_j \left(m_{\alpha }+m_{\beta }\right)\left(m_{\alpha }^2 m_{\beta }+\left(m_{\beta }-6 m_{\alpha }\right)m_{W'}^2\right)+m_{\alpha }^3 m_{\beta }^2-2 \left(2 m_{\alpha}+m_{\beta }\right) m_{W'}^4
\nonumber \\ &&
+m_{\alpha } \left(2 m_{\alpha }+m_{\beta}\right) \left(2 m_{\alpha }+3 m_{\beta }\right) m_{W'}^2\Big)B_0(m_\alpha^2,m_j^2,m_{W'}^2)
+m_{\alpha}\Big(m_j^4 \left(m_{\alpha }+2m_{\beta }\right)+m_j^3 m_{\alpha } \left(m_{\alpha }+m_{\beta}\right)
\nonumber \\ &&
+m_j^2 \left(\left(m_{\alpha }+2 m_{\beta }\right)m_{W'}^2-m_{\beta } \left(m_{\alpha } m_{\beta }+m_{\alpha }^2+2m_{\beta }^2\right)\right)-m_j \left(m_{\alpha }+m_{\beta }\right)\left(m_{\alpha } m_{\beta }^2+\left(m_{\alpha }-6 m_{\beta }\right)m_{W'}^2\right)
\\ \nonumber &&
+m_{\alpha }^2 m_{\beta }^3-2 \left(m_{\alpha }+2m_{\beta }\right) m_{W'}^4+m_{\beta } \left(m_{\alpha }+2 m_{\beta}\right) \left(3 m_{\alpha }+2 m_{\beta }\right) m_{W'}^2\Big)B_0(m_\beta^2,m_j^2,m_{W'}^2)
\nonumber \\ &&
+2m_\alpha m_\beta m_{W'}^2 \left(m^2_{\alpha }-m^2_{\beta }\right) \Big(-3 m_j \left(m_{\alpha}+m_{\beta }\right)+m_j^2-3 m_{\alpha } m_{\beta }-2 m_{\alpha }^2-2m_{\beta }^2
\nonumber \\ &&
+2 m_{W'}^2\Big)C_0(m_\alpha^2,m_\beta^2,0,m_{W'}^2,m_j^2,m_{W'}^2)
+m_\alpha m_\beta\left(m^2_{\alpha }-m^2_{\beta }\right) \left(\left(m_j+m_{\alpha}\right) \left(m_j+m_{\beta }\right)+2 m_{W'}^2\right)
\Big\},
\end{eqnarray}
\begin{eqnarray}
M^{\rm V}_{j,\alpha\beta}&=&\frac{1}{(16\pi)^2m_{\alpha } m_{\beta}(m_\alpha^2-m_\beta^2) m_{W'}^2}\Big\{\left(m_\alpha^2-m_\beta^2\right)\left(m_{W'}^2-m_j^2\right) \Big(-m_j \left(m_{\alpha }+m_{\beta}\right)+m_j^2+m_{\alpha } m_{\beta }
\nonumber \\ &&
+2 m_{W'}^2\Big)B_0(0,m_j^2,m_{W'}^2)-m_\beta\Big(m_j^4 \left(2 m_{\alpha}+m_{\beta }\right)-m_j^3 m_{\beta } \left(m_{\alpha }+m_{\beta}\right)+m_j^2 \Big(\left(2 m_{\alpha }+m_{\beta }\right)m_{W'}^2
\nonumber \\ &&
-m_{\alpha } \left(m_{\alpha } m_{\beta }+2 m_{\alpha}^2+m_{\beta }^2\right)\Big)+m_j \left(m_{\alpha }+m_{\beta }\right)\left(m_{\alpha }^2 m_{\beta }+\left(m_{\beta }-6 m_{\alpha }\right)m_{W'}^2\right)+m_{\alpha }^3 m_{\beta }^2
\nonumber \\ &&
-2 \left(2 m_{\alpha}+m_{\beta }\right) m_{W'}^4+m_{\alpha } \left(2 m_{\alpha }+m_{\beta}\right) \left(2 m_{\alpha }+3 m_{\beta }\right) m_{W'}^2\Big)B_0(m_\alpha^2,m_j^2,m_{W'}^2)
\nonumber \\ &&
+m_\alpha\Big(m_j^4 \left(m_{\alpha }+2m_{\beta }\right)-m_j^3 m_{\alpha } \left(m_{\alpha }+m_{\beta}\right)+m_j^2 \left(\left(m_{\alpha }+2 m_{\beta }\right)m_{W'}^2-m_{\beta } \left(m_{\alpha } m_{\beta }+m_{\alpha }^2+2m_{\beta }^2\right)\right)
\nonumber \\ &&
+m_j \left(m_{\alpha }+m_{\beta }\right)\left(m_{\alpha } m_{\beta }^2+\left(m_{\alpha }-6 m_{\beta }\right)m_{W'}^2\right)+m_{\alpha }^2 m_{\beta }^3-2 \left(m_{\alpha }+2m_{\beta }\right) m_{W'}^4
\nonumber \\ &&
+m_{\beta } \left(m_{\alpha }+2 m_{\beta}\right) \left(3 m_{\alpha }+2 m_{\beta }\right) m_{W'}^2\Big)B_0(m_\beta^2,m_j^2,m_{W'}^2)
+2m_\alpha m_{\beta } m_{W'}^2(m^2_{\alpha}-m^2_{\beta})\Big(3 m_j \left(m_{\alpha}+m_{\beta }\right)
\nonumber \\ &&
+m_j^2-3 m_{\alpha } m_{\beta }-2 m_{\alpha }^2-2m_{\beta }^2
+2 m_{W'}^2\Big)
C_0(m_\alpha^2,m_\beta^2,0,m_{W'}^2,m_j^2,m_{W'}^2)
\nonumber \\ &&
+m_\alpha m_{\beta } (m^2_{\alpha}-m^2_{\beta})\big( -m_j \left(m_{\alpha}+m_{\beta }\right)+m_j^2+m_{\alpha } m_{\beta }
+2 m_{W'}^2 \big)
\Big\},
\end{eqnarray}
\begin{eqnarray}
D_{j,\alpha\beta}&=&\frac{1}{(16\pi)^2 m_{\alpha }m_{\beta} \left(m_{\alpha }-m_{\beta }\right)^2(m_\alpha+m_\beta) m_{W'}^2}
\Big\{
(m_\alpha^2-m_\beta^2)\left(m_{W'}^2-m_j^2\right) \Big(\left(m_j+m_{\alpha }\right)\left(m_j-m_{\beta }\right)
\nonumber \\ &&
+2 m_{W'}^2\Big)B_0(0,m_j^2,m_{W'}^2)+m_\beta\Big(m_j^4 \left(2 m_{\alpha }-m_{\beta }\right)+m_j^3 m_{\beta }\left(m_{\beta }-m_{\alpha }\right)+m_j^2 \Big(\left(2 m_{\alpha}-m_{\beta }\right) m_{W'}^2
\nonumber \\ &&
-m_{\alpha } \left(-m_{\alpha } m_{\beta }+2m_{\alpha }^2+m_{\beta }^2\right)\Big)+m_j \left(m_{\alpha }-m_{\beta}\right) \left(m_{\alpha }^2 m_{\beta }+\left(6 m_{\alpha }+m_{\beta}\right) m_{W'}^2\right)+m_{\alpha }^3 m_{\beta }^2
\nonumber \\ &&
+2 \left(m_{\beta }-2m_{\alpha }\right) m_{W'}^4+m_{\alpha } \left(-8 m_{\alpha } m_{\beta}+4 m_{\alpha }^2+3 m_{\beta }^2\right) m_{W'}^2\Big)B_0(m_\alpha^2,m_j^2,m_{W'}^2)
+m_\alpha\Big(m_j^4 \left(m_{\alpha }-2m_{\beta }\right)
\nonumber \\ &&
+m_j^3 m_{\alpha } \left(m_{\alpha }-m_{\beta}\right)+m_j^2 \left(m_{\beta } \left(-m_{\alpha } m_{\beta }+m_{\alpha}^2+2 m_{\beta }^2\right)+\left(m_{\alpha }-2 m_{\beta }\right)m_{W'}^2\right)
-m_j \left(m_{\alpha }-m_{\beta }\right) \Big(m_{\alpha} m_{\beta }^2
\nonumber \\ &&
+\left(m_{\alpha }+6 m_{\beta }\right)m_{W'}^2\Big)-m_{\alpha }^2 m_{\beta }^3-2 \left(m_{\alpha }-2m_{\beta }\right) m_{W'}^4
\nonumber \\ &&
-m_{\beta } \left(m_{\alpha }-2 m_{\beta}\right) \left(3 m_{\alpha }-2 m_{\beta }\right) m_{W'}^2\Big)B_0(m_\beta^2,m_j^2,m_{W'}^2)
-2m_\alpha m_\beta m_{W'}^2(m_\alpha^2-m_\beta^2)\Big(3 m_{\beta} \left(m_j+m_{\alpha }\right)
\nonumber \\ &&
-3 m_j m_{\alpha }+m_j^2-2 m_{\alpha }^2-2m_{\beta }^2+2 m_{W'}^2\Big)C_0(m_\alpha^2,m_\beta^2,0,m_{W'}^2,m_j^2,m_{W'}^2)
\nonumber \\ &&
-m_\alpha m_\beta(m_\alpha^2-m_\beta^2)\big(\left(m_j+m_{\alpha }\right) \left(m_j-m_{\beta}\right)+2 m_{W'}^2\big)
\Big\},
\end{eqnarray}
\begin{eqnarray}
\overline{D}_{j,\alpha\beta}&=&\frac{1}{(16\pi)^2 m_{\alpha }m_{\beta } \left(m_{\alpha }-m_{\beta }\right)^2(m_\alpha+m_\beta) m_{W'}^2}
\Big\{
(m_\alpha^2-m_\beta^2)\left(m_{W'}^2-m_j^2\right) \Big(\left(m_j-m_{\alpha }\right)\left(m_j+m_{\beta }\right)
\nonumber \\ &&
+2 m_{W'}^2\Big)B_0(0,m_j^2,m_{W'}^2)+m_\beta\Big(m_j^4 \left(2 m_{\alpha }-m_{\beta }\right)+m_j^3 m_{\beta }\left(m_{\alpha }-m_{\beta }\right)
\nonumber \\ &&
+m_j^2 \left(\left(2 m_{\alpha}-m_{\beta }\right) m_{W'}^2
-m_{\alpha } \left(-m_{\alpha } m_{\beta }+2m_{\alpha }^2+m_{\beta }^2\right)\right)-m_j \left(m_{\alpha }-m_{\beta}\right) \left(m_{\alpha }^2 m_{\beta }+\left(6 m_{\alpha }+m_{\beta}\right) m_{W'}^2\right)
\nonumber \\ &&
+m_{\alpha }^3 m_{\beta }^2+2 \left(m_{\beta }-2m_{\alpha }\right) m_{W'}^4+m_{\alpha } \left(-8 m_{\alpha } m_{\beta}+4 m_{\alpha }^2+3 m_{\beta }^2\right) m_{W'}^2\Big)B_0(m_\alpha^2,m_j^2,m_{W'}^2)
\nonumber \\ &&
+m_\alpha\Big(m_j^4 \left(m_{\alpha }-2m_{\beta }\right)+m_j^3 m_{\alpha } \left(m_{\beta }-m_{\alpha}\right)+m_j^2 \left(m_{\beta } \left(-m_{\alpha } m_{\beta }+m_{\alpha}^2+2 m_{\beta }^2\right)+\left(m_{\alpha }-2 m_{\beta }\right)m_{W'}^2\right)
\nonumber \\ &&
+m_j \left(m_{\alpha }-m_{\beta }\right) \left(m_{\alpha} m_{\beta }^2+\left(m_{\alpha }+6 m_{\beta }\right)m_{W'}^2\right)-m_{\alpha }^2 m_{\beta }^3-2 \left(m_{\alpha }-2m_{\beta }\right) m_{W'}^4
\nonumber \\ &&
-m_{\beta } \big(m_{\alpha }-
2 m_{\beta}\big) \left(3 m_{\alpha }-2 m_{\beta }\right) m_{W'}^2\Big)B_0(m_\beta^2,m_j^2,m_{W'}^2)
-2m_\alpha m_\beta m_{W'}^2(m_\alpha^2-m_\beta^2)\Big(3 m_j\left(m_{\alpha }-m_{\beta }\right)
\nonumber \\ &&
+m_j^2+3 m_{\alpha } m_{\beta }-2m_{\alpha }^2-2 m_{\beta }^2+2 m_{W'}^2\Big)C_0(m_\alpha^2,m_\beta^2,0,m_{W'}^2,m_j^2,m_{W'}^2)
\nonumber \\ &&
-m_\alpha m_\beta(m_\alpha^2-m_\beta^2)\big(\left(m_j-m_{\alpha }\right)\left(m_j+m_{\beta }\right)+2 m_{W'}^2\big)
\Big\}.
\end{eqnarray}
\end{widetext}

\section{$\boldsymbol{\Lambda}$-independent factors in leading contributions to transition moments}
\label{anex1}
The factors defining Eqs.~(\ref{MAkj}) to (\ref{Dbarkj}) have the following expressions:
\begin{widetext}
\begin{eqnarray}
\eta^{(1)}_{j}&=&\frac{4 \kappa _j^6-15 \kappa _j^4+12 \kappa _j^2+6 \kappa _j^2 \log\kappa _j^2-1}{(16\pi)^2 \kappa _j^2\left(\kappa _j^2-1\right){}^3},
\\ \nonumber \\
\eta^{(2)}_{j}&=&\frac{2 \kappa _j^8-27 \kappa _j^6+32 \kappa _j^4-9 \kappa _j^2+2 \left(4\kappa _j^4+6 \kappa _j^2-1\right) \kappa _j^2 \log\kappa _j^2+2}{2(16\pi)^2 \kappa _j^2\left(\kappa _j^2-1\right){}^4},
\\ \nonumber \\
\eta^{(3)}_{j}&=&\frac{6 \kappa _j^6-29 \kappa _j^4+26 \kappa _j^2+2 \left(2 \kappa_j^2+5\right) \kappa _j^2 \log\kappa _j^2-3}{2(16\pi)^2 \kappa _j^2 \left(\kappa _j^2-1\right){}^3},
\end{eqnarray}
\begin{eqnarray}
\omega^{(1)}_{j}&=&\frac{-4 \kappa _j^6+15 \kappa _j^4-12 \kappa _j^2-6 \kappa _j^2 \log\kappa _j^2+1}{(16\pi)^2 \kappa _j^2\left(\kappa _j^2-1\right){}^3},
\\ \nonumber \\
\omega^{(2)}_{j,\alpha\beta}&=&\frac{-2 \kappa _j^8+27 \kappa _j^6-32 \kappa _j^4+9 \kappa _j^2-2 \left(4\kappa _j^4+6 \kappa _j^2-1\right) \kappa _j^2 \log\kappa _j^2-2}{2(16\pi)^2 \kappa _j^2\left(\kappa _j^2-1\right){}^4},
\\ \nonumber \\
\omega^{(3)}_{j,\alpha\beta}&=&\frac{6 \kappa _j^6-29 \kappa _j^4+26 \kappa _j^2+2 \left(2 \kappa_j^2+5\right) \kappa _j^2 \log\kappa _j^2-3}{2(16\pi)^2 \kappa _j^2 \left(\kappa _j^2-1\right){}^3}.
\end{eqnarray}
\end{widetext}

\end{document}